\documentclass{article}

\usepackage{geometry}
\usepackage{amsmath}
\usepackage{amssymb}
\usepackage{float}
\usepackage{hyperref}
\usepackage{rotating} 
\usepackage{booktabs}
\usepackage{microtype} 

\usepackage{natbib}
\usepackage{url} 

\begin{document}

\title{Finding your feet: A Gaussian process model for estimating the abilities of batsmen in Test cricket}
\author{Oliver G. Stevenson$^{1,}$\footnote{To whom correspondence should be
addressed. \tt o.stevenson@auckland.ac.nz} \\
\vspace*{0.5em}
Brendon J. Brewer$^1$\\
\vspace*{0.5em}
{\small $^1$Department of Statistics, The University of Auckland}\\[-0.7em]
{\small Private Bag 92019, Auckland 1142, New Zealand}
}

\date{}

\maketitle

\begin{abstract}
In the sport of cricket, player batting ability is traditionally measured using the batting average.
However, the batting average fails to measure both short-term changes in ability that occur during an innings, and long-term changes that occur between innings, due to the likes of age and experience in various match conditions.
We derive and fit a Bayesian parametric model that employs a Gaussian process to measure and predict how the batting abilities of cricket players vary and fluctuate over the course of entire playing careers.
The results allow us to better quantify and predict player batting ability, compared with both traditional cricket statistics, such as the batting average, and more complex models, such as the official International Cricket Council ratings.
\end{abstract}


\setlength{\parskip}{0.0em}

\section{Introduction}

A sportsperson's career is typically full of highs and lows, even for those who are among the world's best at their chosen discipline.
As such, for many sports, the question `who is the best player in the world right now?' can be subjective and difficult to answer.
Often, there is no single, correct answer to such a question, with a wide range of opinions holding some legitimacy, even in a statistically oriented sport like cricket.
Due to the nature of the sport, the majority of professional cricket players routinely experience highs and lows in terms of performance.
Consequently, it can be difficult to quantify the differences in ability between a great player currently experiencing a run of poor performances, and an average player, currently experiencing a run of strong performances.
However, as the sheer volume of data recorded during each match continues to grow, we are slowly able to turn to statistical tools to help us reach data-driven answers to such questions.

With the modern growth in sports analytics, using numeric data to gain an advantage is not a new concept \citep{santos2019}.
The mainstream popularisation and usage of advanced statistics in modern day sports began with `sabermetrics' in baseball \citep{lewis2004moneyball}, and has since had a catalytic influence in the spread of sports analytics globally.
Furthermore, the popularity of data analysis in many sports has overflown from behind the closed doors of coaches' strategy sessions, to public broadcasts and live coverage of matches.
In some sports, advanced statistical measures initially purposed for use by analysts, coaches, and players, have become the gold standard for comparing players amongst more casual fans.
For example, wins above replacement (WAR) in baseball, and both value over replacement player (VORP) and true shooting percentage (TS\%) in basketball, are metrics that were first used by high performance analysts before trickling down to the public.
While the derivation of some of these measures is far from straightforward, fans both with and without formal statistical training are able to digest and interpret the implications these metrics have in regards to their favourite teams and players.

Cricket is no exception; as a sport with a strong statistical culture, steeped in a number of record-keeping traditions (the first known recorded scorecards date as far back as 1776), it has been the subject of numerous academic studies.
However, past research has tended to focus on issues such as \textit{achieving a fair result in an interrupted match} \citep{carter2004, duckworthlewis1998, ian2002, jayadevan2002, stern2016duckworth}, \textit{match outcome prediction} \citep{bailey2006, brooker2011,  brooks2002, davis2015, swartz2009}, and \textit{optimising playing strategies} \citep{clarke1998, clarke1999, clarke2003, norman2010, preston2000, scarf2011, swartz2006}, with less attention paid to developing statistical methods that aim to measure and predict future individual and team performances.

Therefore, given the relative infancy of cricket analytics in the public domain, few advanced metrics for measuring and predicting player performance have been developed.
Moreover, many measures that have been developed tend to suffer from either a lack of a natural cricketing interpretation, or are simply a collection of summary statistics, which while somewhat useful, lack the ability to explain deeper aspects of the game.
Additionally, there has been a recent shift in terms of both funding and mainstream viewership to shorter form `Twenty20' cricket.
As a result, few of these more recently derived metrics have been, or can be, applied to longer form domestic `first-class' and international `Test' cricket.

In this paper, we propose a model for estimating how the batting abilities of individual cricket players vary and fluctuate over the course of entire playing careers, including both individual and match-specific effects. Our methods account for the fact that actual sequences of scores tend to be very noisy, and provide justified uncertainties on the outputs.
Given the recent attention on shorter form cricket and the number of situational complications it brings \citep{davis2015}, our focus is on first-class and Test match cricket, where batting decisions depend less on the match situation.
For as long as statistical records have existed in cricket, the primary measure of an individual's batting ability has been the \textit{batting average}.
The batting average is simply an individual's number of runs scored per dismissal and is an indication of the player's underlying batting ability; generally speaking, the higher the batting average, the better the player is at batting.

However, since it is a single point estimate, the batting average can be interpreted as implicitly assuming that a batsman plays with some constant ability at all times and fails to inform us about variations in a player's batting ability on two scales:
(1) short-term changes in ability, during or within an innings, due to factors such as adapting to the local pitch and weather conditions (a process commonly referred to as `getting your eye in' within the cricketing community); and (2) longer-term changes in ability that are observed between innings, over the course of entire playing careers, due to the likes of age; gaining experience in a variety of match conditions; and general improvements or deteriorations in technique, fitness, or eyesight.

\subsection{Modelling short-term changes in batting ability}
The earliest published statistical study on cricket looked to address the idea of getting your eye in and provided empirical evidence to support the claim that a batsman's scores could be modelled using a geometric distribution, suggesting players do bat with some relatively constant ability during an innings, as implied by the batting average \citep{elderton1945}.
However, it has since been shown that the geometric assumptions do not hold for many players, due to the inflated number of scores of zero (`ducks') that are present in many players' career records \citep{kimber1993, bracewell2009}.

Instead of modelling batting scores directly, \cite{kimber1993} used a nonparametric hazard function to measure how dismissal probabilities change with a batsman's score.
Estimating a batsman's hazard function, $H(x)$, which represents the probability of being dismissed on score $x$, allows us to observe how a player's ability varies during an innings as they score more runs.
The authors found that batsmen were more likely to get out on low scores, early in their innings, than on higher scores, suggesting that batsmen do not bat with some constant ability during an innings, supporting the concept of getting your eye in.
However, due to the sparsity of data at higher scores for many players, these estimates can quickly become unreliable, with estimated dismissal probabilities jumping haphazardly between scores.
\cite{cai2002} addressed this issue by using a parametric smoother on the empirical dismissal probabilities, although given the underlying hazard function used to measure the probabilities is still a nonparametric estimator, the problem of data sparsity remains an issue and continues to distort the estimated hazard function at higher scores.

Motivated by the concept of getting your eye in, \cite{brewer2008} and \cite{stevenson2017} proposed an alternative means of measuring how player batting ability varies during an innings.
Unlike previous studies, a Bayesian parametric model was used to model the hazard function, allowing for a smooth transition in estimated dismissal probabilities between scores, which in the context of batting in cricket, are more realistic than the erratic jumps observed in \cite{kimber1993} and to a lesser extent \cite{cai2002}.

In order to quantify batting ability \emph{during} an innings, \cite{brewer2008} and \cite{stevenson2017} introduced the concept of the \emph{effective batting average}, $\mu(x)$, which represents a batsman's ability on score $x$, in the same units as a batting average.
Given the prevalence of the batting average in cricket, it is far more intuitive for players and coaches to think of ability in terms of batting averages, rather than dismissal probabilities.
In order to model how a player's effective average changes with the number of runs scored, three parameters were fitted to measure (1) a player's initial batting ability when first beginning an innings; (2) a player's equilibrium batting ability once they have their eye in and have adapted to the specific match conditions; and (3) a timescale parameter, measuring how long it takes a batsman to transition from their initial batting state, to their equilibrium batting state.
For the vast majority of past and present Test players analysed, \cite{stevenson2017} found overwhelming evidence to suggest that players are more likely to get out early in their innings, with batting ability improving as they score runs, further supporting the notion of getting your eye in.
The model proposed in \cite{stevenson2017} provides the starting point for the present study.

In a related analysis, \cite{stevenson2017masters} applied a similar model that allowed for score-based deviations in batting ability at any score, not just at the beginning of an innings when a batsman is getting their eye in.
The aim of this study was to determine whether there is any evidence to suggest that batsmen appear to be more likely to get out on certain scores --- as suggested by the cricketing superstition `the nervous 90s' --- whereby players are proposed to bat with an inferior ability due to nerves that may arise when nearing the significant milestone score of 100.
Although there is plenty of anecdotal evidence to suggest that players can get nervous in the 90s, \cite{stevenson2017masters} found no conclusive evidence to suggest batting ability is affected by these nerves.
Instead, some evidence was found to suggest that players are more likely to get out immediately \emph{after} passing significant milestones such as 50 and 100, suggesting that perhaps the `fallible 50s' and `hazardous 100s' would be more justified clich\'{e}s.

\subsection{Modelling long-term changes in batting ability}
While there is plenty of evidence to suggest that players do not bat with some constant ability during an innings, it is also unlikely that a player bats with constant ability throughout an entire playing career.
Instead, variations in a player's underlying ability are likely to occur between innings, due to factors such as how well a player has been performing recently, commonly referred to as `form' in cricket.
It is worth noting that due to the nature of the sport and the process of getting your eye in, even the best batsmen are more likely to fail than succeed on any given day, so the probability distribution over scores is relatively heavy-tailed.
Consequently, it is not uncommon to see players string together a number of low scores in a row, even if their underlying ability has not changed.

Furthermore, there is plenty of anecdotal evidence to suggest that a player's current form can impact mood, anxiety and stress, which in turn can affect performance \citep{totterdell1999, sahni2017}.
Therefore, if recent form has some sort of effect on a player's present batting ability, then the assumption made by numerous studies that each innings in a player's career record can be treated as independent and identically distributed, may not be entirely valid.
It is worth noting that similar concepts have been studied in other sports and have largely reached sceptical conclusions about such effects, for example the concept of the `hot hand' in basketball \citep{gilovich1985, tversky1989}.
Interestingly, more recent studies have suggested that players with a hot hand tend to start taking more difficult shots, which may mask any hot hand effect due to recent form, even if players are making these tough shots at a higher rate than usual \citep{csapo2015}.

If batting form were to have a significant impact on player performance, we should be able to identify periods of players' careers with sequences of high scores (indicating the player was `in form') and sequences of low scores (indicating the player was `out of form') that are statistically significant, loosely speaking.
On the contrary, \cite{durbach2007} found little empirical evidence to support this idea based on the analysis of 16 Test match batsmen.
Instead, the authors concluded that public perceptions of batting form tend to be overestimated, possibly due to a case of recency bias, with the majority of the analysed players' scores able to be modelled using an independent and identically distributed sequence.
Likewise, \cite{kimber1993} justified the use of a batting average to quantify ability as they found no major evidence suggesting the presence of autocorrelation in a player's career record. The model proposed in the present paper allows us to investigate this question in further detail.

In the context of one-day international cricket, \cite{koulis2014} adopted a Bayesian approach that employed a hidden Markov model to determine whether a batsman is in form, or out of form.
The \citet{koulis2014} model defines a number, $K$, of `underlying batting states' for each player, and specifies the expected number of runs to be scored when in each of the $K$ states.
Parameters that measure: \textit{availability} (the probability a batsman is in form for a given match), \textit{reliability} (the probability a batsman is in form for the next $n$ matches) and \textit{mean time to failure} (the expected number of innings a batsman will play before they are out of form), were also estimated for each batsman.
While this approach does suggest that individual batsmen do appear to undergo periods of being in and out of form, a major drawback is that the model requires an explicit and discrete specification of what constitutes an out of form state.
The authors specify a batting state that has a posterior expected median number of runs scored of less than 25, as being out of form, and all other states as being in form.
While in the context of one-day or Twenty20 cricket this is not necessarily an unreasonable specification, there are numerous arguments that could be made to justify a low score scored at a high strike rate, as a successful innings.

Turning our focus to Test match cricket, if a casual fan wanted to determine who the best batsman in the world is, at any point in time, they would have two realistic options.
Firstly, one could rank every active player by their career batting average, and assume the player with the highest average is the best.
While this may give a general indication of ability, there are plenty of instances where the highest career average belongs to a young player, who has only played a handful of matches (and therefore it can be difficult to gauge their true ability), or an older player, who while still a good batsman, is far from the player they once were.
The batting average does not account for the potential impacts of recent player form and does not necessarily relate directly to a player's current ability.

A second approach would be to use the official International Cricket Council (ICC) ratings.
This system utilises a broader range of data to rank players, such as opposition strength, and places more weight on runs scored in more recent innings.
As such, the ICC ratings are generally considered a better indication of current ability than raw batting averages.
However, a look at the current top 10 Test batsmen according to the ICC in Table~\ref{table:iccrankings} illustrates the shortcomings of this approach and raises several questions.  
What does a rating of 911 for Steve Smith tell us about his underlying ability, other than the fact he has the highest ranking of all active players?
What does it mean that Smith is 25 points ahead of Virat Kohli?
This is an example of a rating system, which while useful for ranking players, lacks an intuitive cricketing interpretation and fails to explain the magnitude of differences in ability between players.
Additionally, due to the closed source nature of the ICC rating formula, we do not know how each factor impacts a player's rating, making it difficult to compare with other methods. Finally, it is unclear whether the ICC rating formula attempts to provide inferential or predictive accuracy, or instead tries to formalise expert judgement about who is in and out of form. These two goals may not be entirely compatible.
\begin{table}[h]
	\caption{Official ICC Test batting rankings (as of $1^\text{st}$ December 2020)}
	\centering
	\def\arraystretch{1.2} 
	\begin{tabular}{l l r r}
		\hline
		\textbf{Rank} & \textbf{Player} & \textbf{ICC rating} \\
		\hline
			1. & Steve Smith \hfill (AUS) & 911 \\
			2. & Virat Kohli \hfill (IND) & 886 \\
			3. & Marnus Labuschagne \hfill (AUS) & 827 \\
			4. & Kane Williamson \hfill (NZ) & 812 \\
			5. & Babar Azam \hfill (PAK) & 797 \\
			6. & David Warner \hfill (AUS) & 793 \\
			7. & Cheteshwar Pujara \hfill (IND) & 766 \\
			8. & Ben Stokes \hfill (ENG) & 760 \\
			9. & Joe Root \hfill (ENG) & 738 \\
			10. & Anjinkya Rahane \hfill (IND) & 726 \\
		\hline
	\end{tabular}
	\label{table:iccrankings}
\end{table}

The approach taken in \cite{boys2019} overcomes these issues with the ICC ratings.
In this study, the authors fit a Poisson random effects model to estimate the age at which individual players would reach their peak batting ability.
The model estimates both the age at which individual players reach their peak and their expected batting average at this age, allowing for the comparison of players across eras.
These estimates provide us with an approximation of the differences in ability between players at their peak, in a more meaningful manner than the ICC's rating points.

Our proposed model builds on those detailed in \cite{stevenson2017}, such that we can not only measure and predict how player batting abilities fluctuate in the short-term, during an innings, but also in the long-term, between innings, over the course of entire playing careers.
A preliminary version of this model was presented in \cite{stevenson2018}.
Similar to \cite{boys2019}, the model estimates player ability in terms of an expected number of runs to be scored in an innings and provides a more accurate quantification of an individual's ability at any point of their career than traditional cricket metrics, such as the batting average.
This allows us to treat batting form as continuous rather than binary; instead of defining players as being in or out of form, we can describe them as improving or deteriorating in terms of ability.

Additionally, by estimating the effect that form and recent performances have on each player, we are able to investigate the cricketing concept of `finding your feet', whereby players do not begin their careers playing to the best of their ability.
Rather, it takes time for new players to adjust to the demands of domestic or international cricket, before reaching their peak.
Gaining a deeper understanding of how long it can take an individual player to step up and fulfill their potential would be of great use to both coaches and players.
It is a familiar sight to fans from all countries to see a new player given just a couple of matches for their national side to prove themselves, before being discarded, or conversely, there are numerous players who seem to hang around the international scene well past their use by date, without managing to post an individual performance of note.

\section{Model specification}
\subsection{Model likelihood}
The derivation of the model likelihood {\em for a single innings} follows the method detailed in \cite{brewer2008} and \cite{stevenson2017, stevenson2018}.
If $X \in \{0, 1, 2, 3, ...\}$ is the number of runs a batsman scores in a particular innings, we define a \textit{hazard function}, $H(x) \in [0, 1]$, as the probability a batsman gets out on score $x$, that is, $H(x) = P(X = x | X \geq x)$.
Assuming a functional form for $H(x)$, conditional on some parameters $\theta$, we can calculate the probability distribution for $X$ as follows:
\begin{equation} \label{eq:px}
	P(X = x) = H(x) \prod_{a = 0}^{x - 1} \left[ 1 - H(a) \right].
\end{equation}

For any given value of $x$, this can be thought of as the probability of a batsman surviving up until score $x$, then being dismissed.
When we infer our model parameters, $\theta$, from data, Equation \ref{eq:px} provides the likelihood function for a single innings.
However, in cricket there are a number of instances where a batsman's innings may end without being dismissed (referred to as a `not out' score).
Therefore, in the case of not out scores, we compute $P(X \geq x)$ as the likelihood, rather than $P(X = x)$.
Comparable to right-censored observations in the context of survival analysis, this assumes that for not out scores the batsman would have gone on to score some unobserved score, conditional on their current score and their assumed hazard function.
Treating not out scores in this manner implies that the sequence of out/not out flags in a given dataset, without the associated score, is uninformative about any of our model parameters.

Therefore, if $T$ is the total number of innings a player has batted in and $N$ is the number of not out scores, the probability distribution for a set of conditionally independent `out' scores $\{x_t\}_{t = 1}^{T - N}$ and `not out' scores $\{y_t\}_{t = 1}^N$ can be expressed as 
\begin{equation} \label{eq:pxy}
	p(\{x\}, \{y\}) = \prod_{t = 1}^{T - N} \Big(H(x_t) \prod_{a = 0}^{x_t - 1} [1 - H(a)] \Big) \times \prod_{t = 1}^N \Big(\prod_{a = 0}^{y_t - 1} [1 - H(a)] \Big).
\end{equation}

When data $\{x, y\}$ are fixed and known, Equation \ref{eq:pxy} gives the likelihood for any proposed form of the hazard function, $H(x; \theta)$.
Therefore, conditional on the set of parameters $\theta$ governing the form of $H(x)$, the log-likelihood, $\ell(\theta)$, is
\begin{equation} \label{eq:likelihood}
		\ell(\theta) = \sum_{t = 1}^{T - N} \textup{log} \ H(x_t) + \sum_{t = 1}^{T - N} \sum_{a = 0}^{x_t - 1} \textup{log} [1 - H(a)] + \sum_{t = 1}^N \sum_{a = 0}^{y_t - 1} \textup{log} [1 - H(a)].
\end{equation}

The equations presented in this section define the likelihood function for any
proposed theory for how ability varies as a function of current score. Our
specific parameterisation of this is given in Section~\ref{sec:within_innings},
and is equivalent to the model used by \citet{stevenson2017} for a single innings.
However, the primary difference between this paper and \citet{stevenson2017} is
that we extend the model to allow for both short and long-term time variation
by allowing those parameters to depend on time and other match-specific factors,
as well as on the current score (Section~\ref{sec:betweeneffects}).

\subsection{Within-innings effects}\label{sec:within_innings}
In order to estimate how an individual's batting ability changes during or within an innings, due to the process of getting your eye in, we adopt the same parameterisation of the hazard function as detailed in \cite{stevenson2017}.
Rather than estimating player ability using the hazard function, we introduce the \textit{effective average function}, $\mu(x)$, which represents a player's ability when on score $x$, in units of a batting average. 
The hazard function can then be expressed in terms of the effective average function, $\mu(x)$, using
\begin{equation} \label{eq:hx}
	H(x) = \frac{1}{\mu(x) + 1},
\end{equation}
where the effective average contains three parameters, $\theta = \{\mu_1, \mu_2, L\}$, and takes an exponential functional form as per Equation \ref{eq:mux}.
\begin{equation} \label{eq:mux}
	\mu(x) = \mu_2 + (\mu_1 - \mu_2)\exp \left(-\frac{x}{L}\right)
\end{equation}

Under this specification, $\mu_1$ represents a player's initial batting ability when beginning a new innings, while $\mu_2$ is the player's eye in, peak, or equilibrium batting ability, once used to the specific match conditions.
Both $\mu_1$ and $\mu_2$ are expressed in units of a batting average.
The timescale parameter $L$, measures the speed of transition between $\mu_1$ and $\mu_2$ and is formally the $e$-folding time.
By definition the $e$-folding time, $L$, signifies the number of runs scored for approximately 63\% (formally $1 - \frac{1}{e}$) of the transition between $\mu_1$ and $\mu_2$ to take place and can be understood by analogy with a `half-life'.
By observing the posterior distributions for these parameters, the model is able to quantify how well a batsman is batting at any given stage of an innings and answer questions about individual players, such as (1) how well players perform when they first beginning a new innings, (2) how much better players perform once they have their eye in, and (3) how long it takes them to get their eye in.

\subsection{Between-innings effects} \label{sec:betweeneffects}
At this point, the model is equivalent to that of \citet{stevenson2017} and can estimate how much a player's ability changes during an innings, but assumes that a player's underlying ability is constant throughout his or her entire career.
In reality, our estimates for an individual player's expected scores should fluctuate from innings to innings, due to a number of match and individual-specific factors.

\subsubsection{Individual-specific effects} \label{sec:individualeffects}
To allow the model to account for both short-term, within-innings changes, and long-term, between-innings changes in ability, we extend the effective average function in Equation \ref{eq:mux} by introducing a time dependence, such that
\begin{equation*}
	\mu(x, t; \theta) := \textup{batting ability on score } x \textup{, in } t^{th} \textup{ career innings},
\end{equation*}
where $\mu(x, t; \theta)$ is expressed in units of a batting average, and $t$ indicates the $t^{th}$ innings of a player's career.
By taking the expectation over all scores, $x$, at a given time, we obtain $\nu(t)$, where
\begin{equation*}
	\nu(t; \theta) := \textup{expected number of runs scored in } t^{th} \textup{ career innings},
\end{equation*}
which, conditional on the model parameters, is equivalent to a player's expected batting average in their $t^{th}$ career innings.

If all parameters, $\theta$, determining $\mu(x, t)$ at a particular innings $t$ are known,
$\nu(t)$ is given by a deterministic function of those parameters; it is the
expectation of the implied distribution over $x_t$.
When we estimate $\nu(t)$ for an individual player, we want to account for time variation in ability between innings, due to factors such as recent form, age, experience, and the element of randomness associated with cricket.
This is achieved by fitting a unique $\mu_2$ parameter for each innings in a player's career, where $\mu_{2_t}$ represents a player's eye in batting ability in their $t^{th}$ career innings.

Restricting the set of parameters that are allowed to vary with time to only include $\mu_2$ implies that we are assuming a player gets their eye in at a similar rate in each of their career innings.
From a cricketing perspective, it makes sense that the process of getting your eye in is more closely related to an individual's playing style, which for many players remains relatively constant over their Test careers, rather than their underlying ability.
Conceptually, $\mu_{2_t}$ can be thought of as an innings-specific `skill-ceiling', which we believe is more likely to vary over the course of a career, than a player's initial batting ability, $\mu_1$, or the rate at which they get their eye in, $L$.
Of course, we could allow both $\mu_1$ and $L$ to also vary with time, however, this would require us to introduce a number of new parameters, which given the already large parameter space, would result in severely decreased computational efficiency.

The innings-specific effective average function, $\mu(x, t)$, can then be expressed as
\begin{equation} \label{eq:muxt}
	\mu(x, t) = \mu_{2_t} + (\mu_1 - \mu_{2_t}) \ \exp \left(-\frac{x}{L} \right),
\end{equation}
where the dependence of $\mu_2$ on time, $t$, is explicitly noted.

To allow a player's underlying batting ability a reasonable amount of flexibility to vary between innings, the prior for the set of $\{\mu_{2_t}\}$ terms is specified using a Gaussian process.
A Gaussian process is fully specified by an underlying mean value, $\lambda$, and a covariance function $K(t_j, t_k)$, where $t_j$ represents the index of a player's $j^{th}$ career innings and $t_k$ represents the index of a player's $k^{th}$ career innings.
Therefore, our selected form of the covariance function, $K(t_j, t_k)$, will determine how much a player's batting ability can vary from innings to innings \citep{mackay2003, rasmussenwilliams2006}.

A number of covariance functions are available to choose from, depending on how much we wish to allow player ability to fluctuate between innings.
A common choice is the squared exponential covariance function, which was the choice of covariance function in \cite{stevenson2018}.
However, we have since determined that the squared exponential function does not allow for enough short-term variation in ability between innings since it implies smooth functions with high prior probability.
Therefore, in order to capture variations in ability due to the likes of recent form, age, and experience, we have opted to use a powered exponential covariance function, which contains scale and length parameters, $\sigma$ and $\ell$, as well as a smoothing parameter $\alpha$, which controls the smoothness of the resulting function for $\mu_{2_t}$.
Our chosen covariance function takes the form
\begin{equation} \label{eq:covariance}
	K(t_j, t_k) = \sigma^2 \textup{exp} \left( - \frac{|j - k|^\alpha}{\ell^\alpha} \right),
\end{equation}
where $\alpha \in \left[ 1, 2 \right]$.
As $\alpha \to 2$, this covariance function converges to the smoother squared exponential covariance function, and as $\alpha \to 1$ it converges to an Ornstein-Uhlenbeck process or AR(1) process, which is a special case of the Mat\'ern class of covariance functions.
Therefore, by assessing the posterior distribution of our smoothness parameter, $\alpha$, we can get an idea of how much an individual's ability appears to fluctuate in the shorter term.

\subsubsection{Venue and innings-specific effects}
As well as accounting for effects that may be present at an individual level, we also ought to consider several factors that are specific in the context of each individual match.
A major characteristic that sets cricket apart from many sports, is the significant role the pitch and weather conditions can play in a match.
Pitches differ greatly around the world as a result of local climate, varieties of soil available and preparation techniques.
Vastly different approaches to batting and bowling are seen between the dusty, spin-friendly pitches of the sub-continent; the hard, flat pitches found in Australia; and the greener pitches commonly prepared in England and New Zealand.
No two pitches will play exactly the same and as such it can be difficult to adjust to a new pitch from match to match, particularly when playing away from home, in a foreign environment.

Therefore, due to the varying nature of cricket pitches, it is fair to assume that some players will perform better in their home country, where they are more familiar with the local playing conditions.
Additionally, many sportspeople simply enjoy more success when playing in front of a home crowd \citep{pollard1986, nevill1999}.
However, given the comparatively subdued nature of cricket crowds (particularly in Test cricket) with other sports, any observable home ground effect is more likely to be due to a familiarity with pitch and weather conditions, rather than an effect due to the crowd itself \citep{morley2005}.

Also of note is the manner in which pitches tend to deteriorate during a match, which may also have an effect on a player's performance.
First-class and Test matches can span up to five days with each team batting twice, for a total of four innings in a match.
Therefore, in many countries batting is believed to become more difficult the longer a match goes on, as the condition of the pitch will often deteriorate due to exposure to the elements and general wear and tear.
As stated by the infamous 19$^\text{th}$-century England captain, W. G. Grace:
\newline

\textit{``When you win the toss --- bat. If you are in doubt, think about it, then bat. If you have very big doubts, consult a colleague then bat.''} --- W.G. Grace
\newline
 
While this bat-first philosophy is not strictly the best approach for every single pitch, the data do suggest that batting is generally easier during a team's first innings of a match, compared with their second innings.
To account for such effects, we first introduce two indicator variables, $v_t$ and $i_t$, representing the match venue and whether it is a team's first or second innings of the match, for the player's $t^{th}$ career innings.
\begin{equation}
	v_t=
    \begin{cases}
      1, & \text{if playing at a home venue} \\
      0, & \text{if playing at a neutral venue} \\
      -1, & \text{if playing at an away venue}
    \end{cases}
 \end{equation}
\begin{equation}
	i_t=
    \begin{cases}
      1, & \text{if team's first innings of a match} \\
      -1, & \text{if team's second innings of a match} \\
    \end{cases}
\end{equation}

To estimate the venue and innings-specific effects, we introduce two new model parameters, $\psi$ and $\phi$, to our innings-specific effective average function, $\mu(x, t)$, such that

\begin{equation} \label{eq:muxt effects}
	\mu(x, t) = \left[\mu_{2_t} + (\mu_1 - \mu_{2_t}) \ \textup{exp} \left( \frac{-x}{L} \right) \right] \times \psi^{v_t} \times \phi^{i_t}.
\end{equation}

Our venue variable, $v_t$, has three levels, with the baseline level being a neutral venue.
An estimate $\psi > 1$ indicates that a player performs better in home conditions, while an estimate $\psi < 1$ indicates a player performs better at venues away from home.
For the innings-specific effect, an estimate $\phi > 1$ indicates that a player tends to perform better in their team's first innings of a match, while an estimate $\phi < 1$ indicates a player performs better in their team's second innings.

We can then obtain an estimate for $\nu(t)$ by taking the expectation over all scores, $x$, which can be computed analytically using Equation \ref{eq:muxt effects}.
Under this model specification, $\nu(t)$ represents a player's underlying ability in units of an expected number of runs to be scored in their $t^{th}$ career innings, assuming the match is at a neutral venue and the innings number in the match is unknown, with a 50/50 prior for whether a team is batting in their first or second innings.

\subsection{Prior distributions and model fitting}
In this section we specify our prior distributions.
As per Section \ref{sec:betweeneffects}, the model contains the following set of parameters, $\theta$.
\begin{equation} \label{eq:theta}
	\theta = \{ \mu_1, \{\mu_{2_t} \}, L, \lambda, \sigma, \ell, \alpha\, \psi, \phi \}
\end{equation}

For ease of prior specification, for the within-innings parameters, $\mu_1$ and $L$, we change coordinates as detailed in \cite{stevenson2017}, introducing parameters $C$ and $D$, with the parameters assigned the following prior distributions.
This specification ensures that $\{\mu_1, L\} < \mu_2$ and implies that we do not expect a batsman to get any worse as they get their eye in.
\begin{equation*}
	\begin{aligned}[c]
  		C \sim \text{Beta}(1, 2); \ & \mu_{1,t}\leftarrow C\mu_{2_t} \\
		D \sim \text{Beta}(1, 5); \ & L_t \leftarrow D\mu_{2_t}
	\end{aligned}
\end{equation*}

As detailed in Section \ref{sec:individualeffects}, our prior for the set of $\{ \mu_{2_t} \}$ parameters is specified using a Gaussian process, with an underlying constant mean value, $\lambda$, and covariance function, $K(t_j, t_k; \sigma, \ell, \alpha)$ from Equation \ref{eq:covariance}.
As we are measuring batting ability in terms of a batting average (which by definition must be positive), we actually model $\text{log}\{ \mu_{2_t} \}$ and back-transform accordingly, rather than modelling $\{ \mu_{2_t} \}$ itself, to ensure positivity of $\mu_{2_t}$ for the entire parameter space.

The model parameters, prior distributions and relevant definitions are summarised in Table \ref{tab:summary}.
These priors are generally informative, but conservative, loosely reflecting our cricketing knowledge.
For example, the lognormal prior for $\lambda$ simply suggests we expect the median player to have an eye in, or peak batting ability of around 25 runs, which in the context of Test cricket is a reasonable assumption.

\begin{table}[h]
	\caption{The hyperparameters, parameters, data and effective average functions, including the prior distribution for each quantity where relevant.}
	\centering
	\resizebox{\textwidth}{!}
   	{
	\def\arraystretch{1.2} 
	\begin{tabular}{l l l}
		\hline
		\textbf{Quantity} & \textbf{Interpretation} & \textbf{Prior} \\
		\hline
		\multicolumn{3}{l}{\textbf{Data}} \\
        \hline
		$t$ & Career innings index (time) & \\
        $o_t$    & Out/not out flag in $t^{th}$ career innings & \\
		$v_t$ & Venue in $t^{th}$ career innings & \\
		$i_t$ & Team innings \# in $t^{th}$ career innings & \\
		$x_t$ & Runs scored in $t^{th}$ career innings & Likelihood function given in Equation \ref{eq:likelihood} \\
		\hline
		\multicolumn{3}{l}{\textbf{Within-innings effects}} \\
		\hline
		$\mu_{1,t}$ & Initial batting ability in $t^{th}$ career innings & $C \sim \text{Beta}(1, 2);  \mu_{1,t}\leftarrow C\mu_{2_t}$ \\
		$L_t$ & Transition parameter in $t^{th}$ career innings & $D \sim \text{Beta}(1, 5); L_t \leftarrow D\mu_{2_t}$ \\
		\hline
		\multicolumn{3}{l}{\textbf{Between-innings effects}} \\
		\hline
		$\{ \mu_{2_t} \}$ & Eye in batting ability in $t^{th}$ career innings & $\textup{log}\{\mu_{2_t}\} \sim \textup{GP}(\lambda, K(t_j, t_k; \sigma, \ell, \alpha))$ \\
		$\lambda$ & Mean value of Gaussian process & $\text{log}(\lambda) \sim \text{Normal}(\textup{log}(25), 0.75^2)$ \\
		$\sigma$ & Scale parameter of covariance function, $K(t_j, t_k)$ & $\text{log}(\sigma) \sim \text{Normal}(\textup{log}(0.2), 1^2)$ \\
		$\ell$ & Length parameter of covariance function, $K(t_j, t_k)$ & $\text{log}(\ell) \sim \text{Normal}(\textup{log}(20), 1^2)$ \\
		$\alpha$ & Smoothing parameter of covariance function, $K(t_j, t_k)$ & $\alpha \sim \text{Uniform}(1, 2)$ \\
		\hline
		\multicolumn{3}{l}{\textbf{Match-specific effects}} \\
		\hline
		$\psi$ & Venue effect & $\log(\psi) \sim \text{Normal}(\log(1), 0.25^2)$ \\
		$\phi$ & Team innings \# effect & $\log(\phi) \sim \text{Normal}(\log(1), 0.25^2)$ \\
		\hline
		\multicolumn{3}{l}{\textbf{Covariance and effective average functions}} \\
		\hline
		$K(t_j, t_k)$ & Covariance function for Gaussian process & Functional form given in Equation \ref{eq:covariance} \\
		$\mu(x, t)$ & Batting ability on score $x$, in $t^{th}$ career innings,& Functional form given in Equation \ref{eq:muxt} \\
		& in units of a batting average \\
		$\nu(t)$ & Expected number of runs scored in $t^{th}$ career innings & Computable from $\mu(x,t)$ \\
		\hline
	\end{tabular}
	}
	\label{tab:summary}
\end{table}

The most restrictive prior is that for $\sigma$, which has been catered specifically for the modelling problem at hand.
The proposed prior for $\sigma$ implies that the median player's underlying eye in batting ability will vary from the underlying mean, $\lambda$, by approximately 20\% over the course of their playing career.
If left unchecked, the model can have a tendency to fit Gaussian processes with large values for $\sigma$ to best fit batting career data that is typically very noisy and can range from having a score of zero in one innings, to a large score of 100 or more in the next.
While many of the proposed Gaussian processes with large values for $\sigma$ may theoretically fit the data better, in many cases it is simply not believable that a player's underlying batting ability will have changed by such a significant amount between two innings.

\section{Data}
\begin{figure}[h]
	\centering
	\includegraphics[width = 0.7\linewidth]{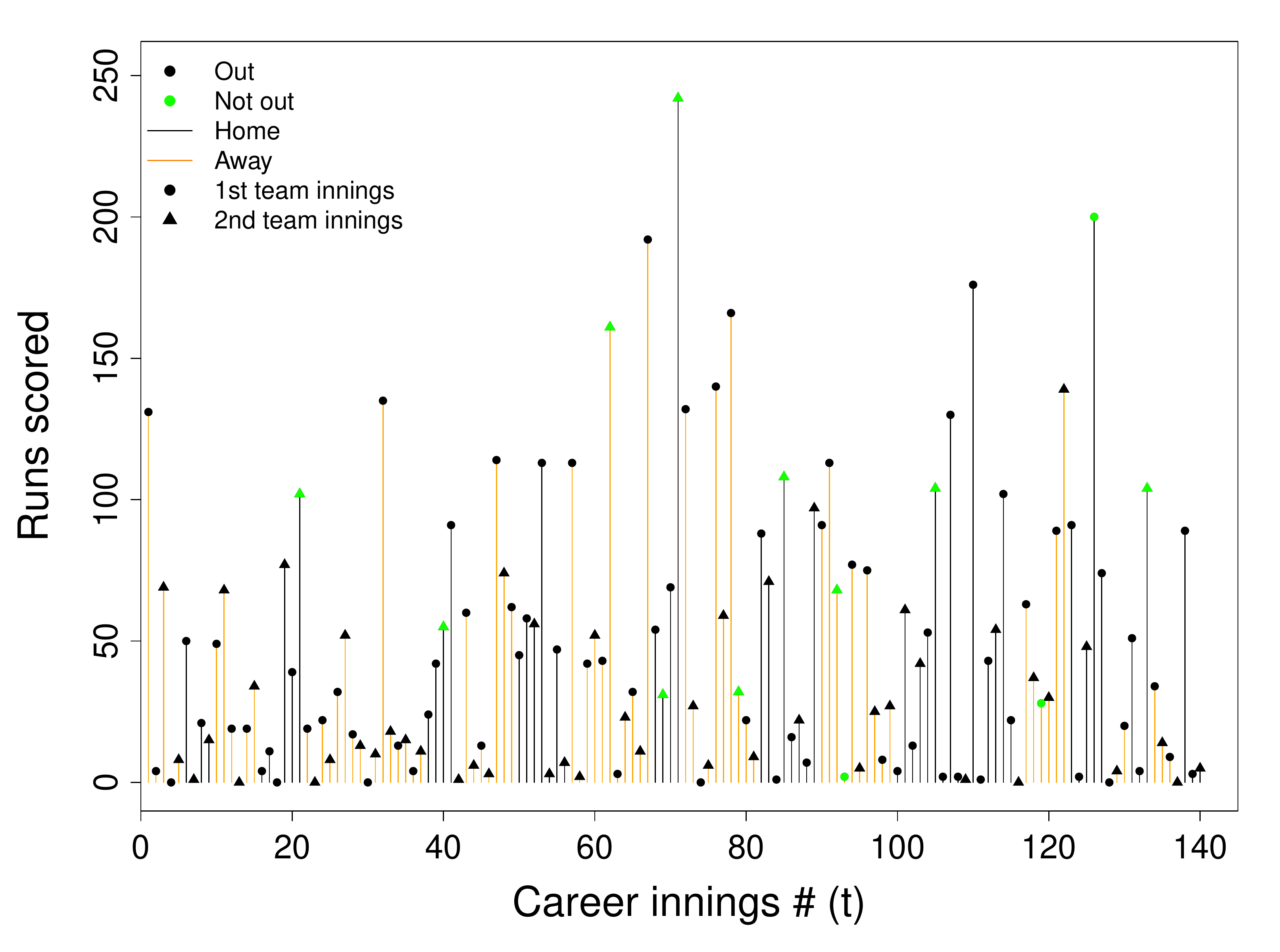}
	\caption{Test match career batting data for current New Zealand captain, Kane Williamson.}
	\label{fig:data}
\end{figure}

The data used to fit the present model are simply the Test career scores of an individual batsman and are obtained from Statsguru, the cricket statistics database on the Cricinfo website (\url{www.espncricinfo.com}).
To visualise an example of the data that is used to fit the model, the Test career scores for current New Zealand Test captain, Kane Williamson, are plotted in Figure \ref{fig:data}.
The data for Williamson illustrates the high innings-to-innings variation in scores, while also showing that batsmen do not always bat twice in a given match.
The latter can be attributed to a variety of causes, such as the match coming to a premature conclusion as a result of bad weather, or because the player's second innings was not required as their team had already won the match.

Due to a combination of law changes and technological advancements, the pace and format of Test cricket has changed considerably since the first Test was played in 1877.
In order to maintain a modern outlook on the game, we have opted to fit the model to the career data of all players who have batted in a Test match innings in the $21^\text{st}$ century.
This corresponds to a total of 1018 players, from 12 different countries, who have batted in a combined total of 40273 innings.

For each innings, the data contain the number of runs scored as well as indicating whether or not the batsman's innings ended as a not out score.
Additional information, such as the venue and whether the innings took place during their team's first or second innings of the match, are also available.
A cursory inspection of the data suggests our belief that players tend to bat better at home, in their first innings of a match, is reasonable.
According to the empirical data, player batting averages are approximately 17\% higher at home venues, compared with away venues, and are approximately 20\% higher when batting in a team's first innings of a match, compared with a team's second innings.
This suggests we are likely to find some evidence of venue and innings-specific effects that we can quantify on an individual basis, via the parameters $\psi$ and $\phi$.

\section{Results}
As the model requires a set of $\{ \mu_{2_t} \}$ terms to be fitted (one for each innings in a player's career), there can be a large number of model parameters for players who have enjoyed long careers.
Therefore, to fit the model, we employed a C++ \citep{c++} implementation of the Nested Sampling (NS) algorithm \citep{skilling2006} that uses Metropolis-Hastings updates to generate the new points.
Nested Sampling is a useful approach in this case, as it is able to handle both high dimensional and multimodal problems that may arise.
For each player analysed, the model output provides us with the posterior distribution for each of the model parameters, as well as the marginal likelihood, which makes for trivial model comparison.
The results reported in this paper are based on NS runs with 1000 particles and 1000 MCMC steps per NS iteration. The results were not sensitive to these values, indicating that the sampling was sufficiently effective.

\subsection{Analysis of individual players}
By drawing posterior samples of the model parameters we are able to compute the posterior predictive distribution for $\nu(t)$, for individual batsmen, illustrating the progression of their Test batting careers to date, as well as forecasting future batting abilities.
In order to better understand the implications of the model, we will focus on the results for `the big four' -- Kane Williamson (New Zealand), Steve Smith (Australia), Virat Kohli (India) and Joe Root (England) -- four players who have widely been regarded as the best batsmen in world cricket, in recent years.
Career trajectories for all 1018 players analysed are available to view in a RShiny application at \url{http://oliverstevenson.co.nz/cricket-visualisation/}.
\begin{figure}[h]
	\centering
	\includegraphics[width = 0.7\linewidth]{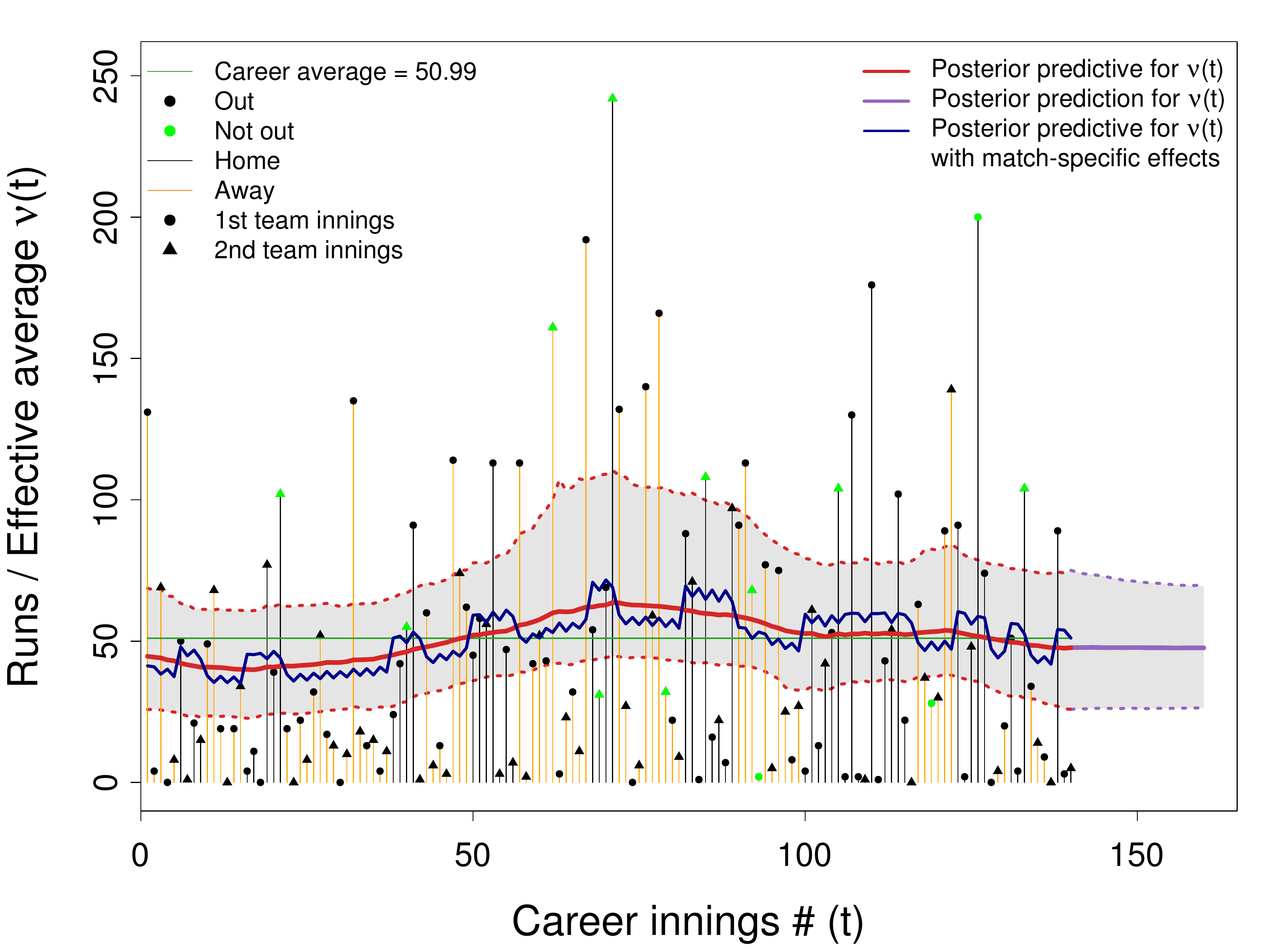}
	\caption{Test match batting career trajectory (i.e. the posterior predictive estimate for $\nu(t)$) for Kane Williamson, including the 95\% credible interval (shaded region) for $\nu(t)$.}
	\label{fig:GPWilliamson}
\end{figure}

The evolution of Kane Williamson's Test match batting career is shown in Figure \ref{fig:GPWilliamson}.
The posterior predictive estimate for $\nu(t)$ (red) represents Williamson's underlying batting ability in units of a batting average, assuming the match was played at a neutral venue and his team's innings number within the match is unknown.
The dark blue line represents the posterior predictive estimate for $\nu(t)$, given that we know the venue and innings-specific variables, $v_t$ and $i_t$.
Each of these posterior estimates are computed by taking the median of a large number of posterior samples and provides an estimate for the expected number of runs to be scored in each career innings.
The posterior median was used rather than the posterior mean, as the posterior distribution for $\nu(t)$ is not necessarily symmetric and can have heavy tails.
Additionally, the posterior median was found to provide more accurate predictions of future scores when computing the prediction error in Section \ref{sec:error}.

\begin{figure}[h]
	\centering
	\includegraphics[width = 0.7\linewidth]{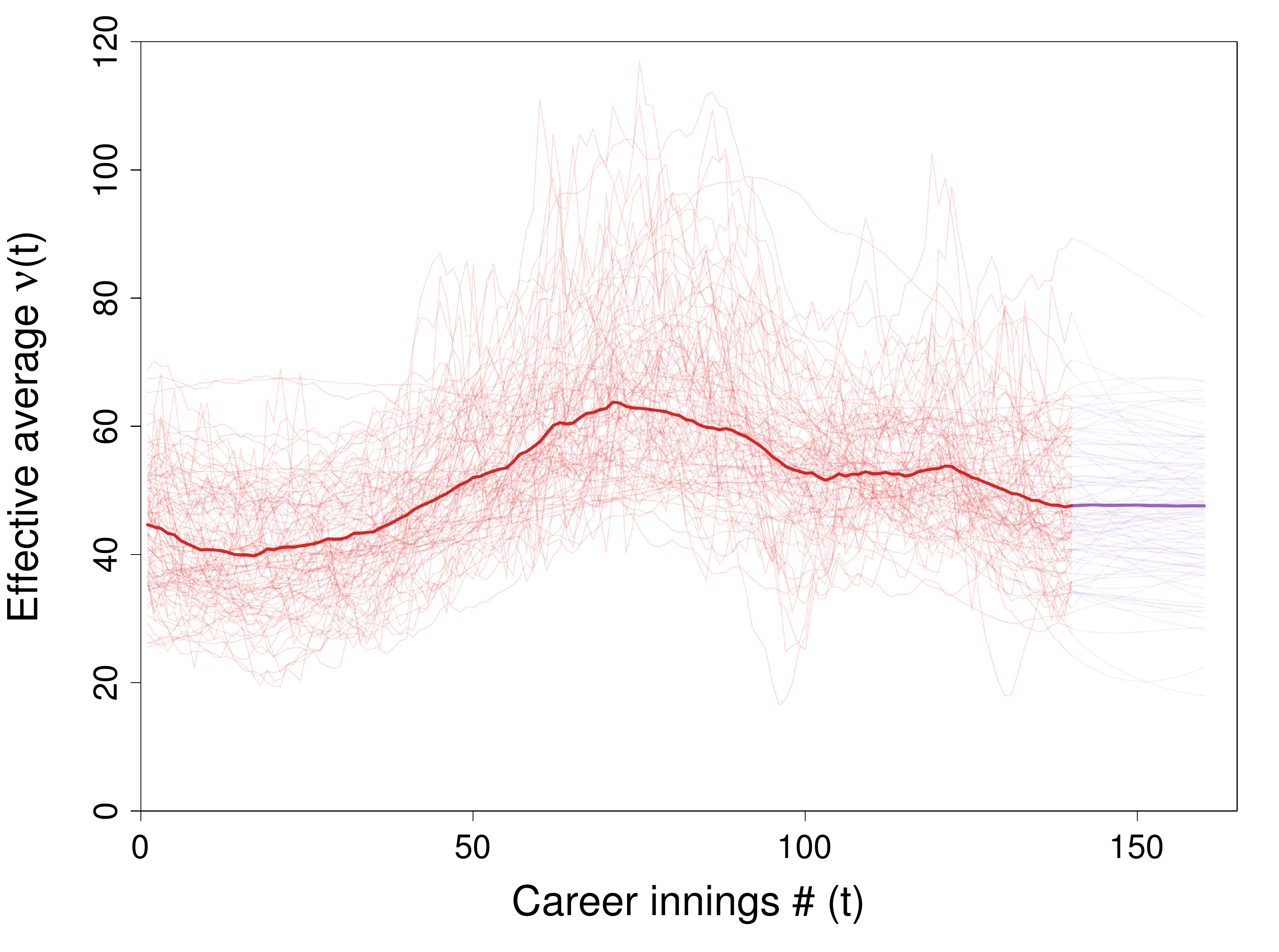}
	\caption{A subset of posterior samples for $\nu(t)$, the expected score given the parameters, for Kane Williamson.
             The purple lines represent predictions for $\nu(t)$ for 20 innings into the future.
             Due to the noisiness of batting scores, a wide variety of ability trajectories are compatible with the data.}
	\label{fig:Williamson_samples}
\end{figure}

A subset of the posterior samples used to compute the posterior predictive estimate for $\nu(t)$ are shown in Figure \ref{fig:Williamson_samples}, while the posterior distributions for each of the model parameters are plotted in Figures \ref{fig:between_posteriors} and \ref{fig:match_posteriors}, alongside the corresponding prior distributions.
Posterior summaries for each of the model parameters are provided in Table \ref{tab:parameters}, allowing us to estimate the magnitude of the various effects in the model.

\subsubsection{Between-innings effects}
As shown in Figure \ref{fig:GPWilliamson} the model estimates that early in his career, Williamson's underlying ability was less than that implied by his current career average of 50.99, as indicated by the comparatively lower estimates for $\nu(t)$ early in his career.
However, over time his underlying ability appears to have improved, likely as a result of gaining experience and exposure against world-class bowling attacks, in a variety of pitch and weather conditions.
The estimates for Williamson's current and future abilities suggest that up until recently, he was a better batsmen than his career average indicates.
If we are to believe in the cricketing concept of `finding your feet', this is not a particularly surprising result, as we should expect players to take a number of innings to adjust to the demands of international cricket before reaching their peak ability.

\begin{figure}[H]
	\centering
	\includegraphics[width = 0.7\linewidth]{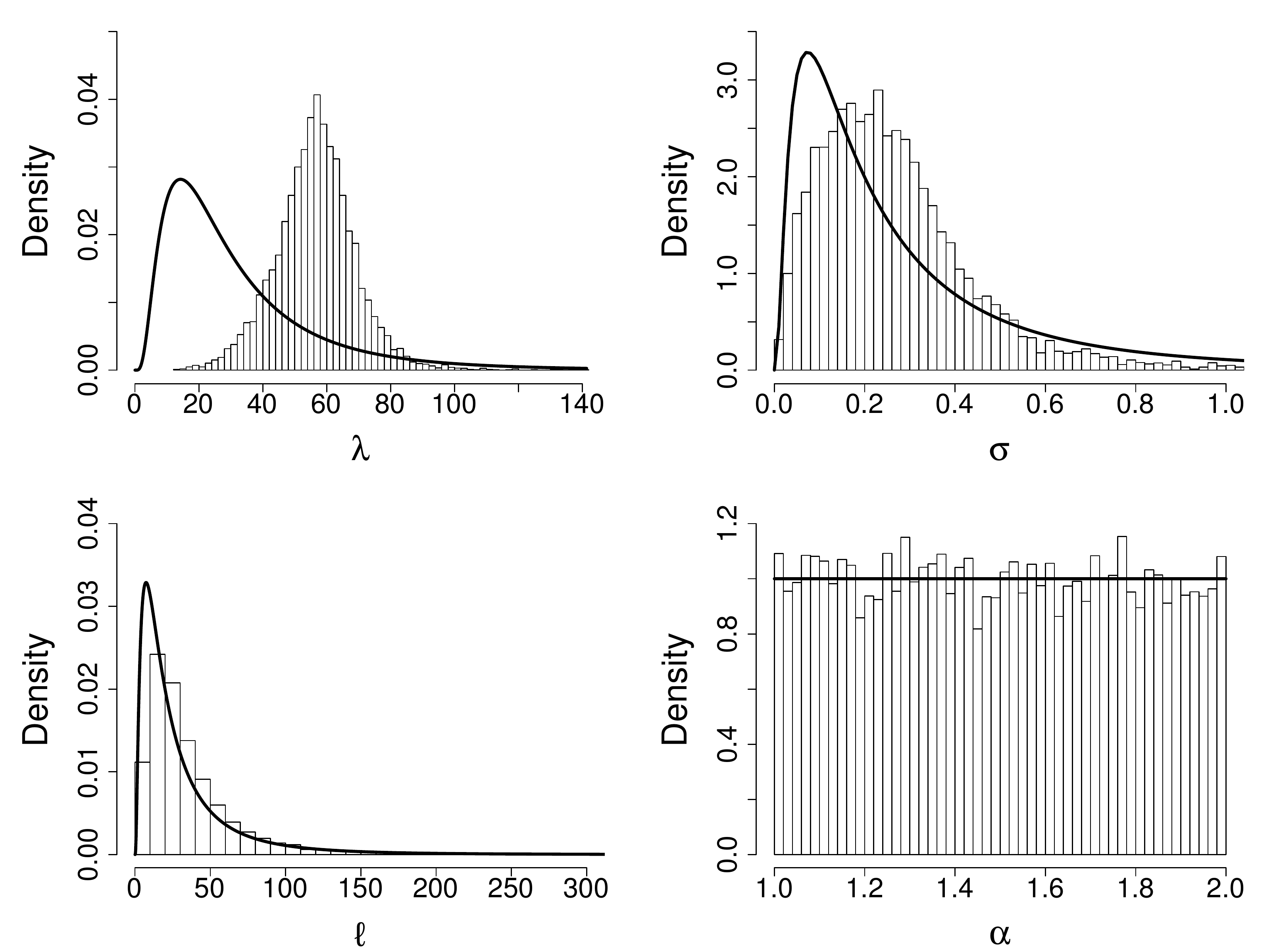}
	\caption{Parameter posterior distributions for the Gaussian process hyperparameters
                $\lambda$, $\sigma$, $\ell$ and $\alpha$ for
                Kane Williamson. Black lines indicate the prior distributions.
                The data are barely able to modify the prior distribution for $\ell$ and
                $\alpha$, suggesting that we cannot distinguish between
                smooth or ragged underlying trajectories because of the
                noisy data.}
	\label{fig:between_posteriors}
\end{figure}

The posterior distributions for the Gaussian process parameters are plotted in Figure \ref{fig:between_posteriors} and suggest that the data for Kane Williamson have been reasonably informative with respect to $\lambda$ and therefore our set of $\{\mu_{2, t}\}$ terms.
The posterior for $\sigma$ has shifted away from zero and is of some significance; there is little posterior weight assigned to values of $\sigma$ close to zero, which provides us with some evidence to support the presence of long-term variation in Williamson's underlying batting ability.
However, the data have not informed us greatly with respect to parameters $\ell$ and $\alpha$, which provides us with little evidence to either confirm or refute the presence of short-term, innings-to-innings variation in ability.

\subsubsection{Venue and innings-specific effects}
If we consider the posterior predictive estimate for $\nu(t)$, including the venue and innings-specific effects (dark blue) in Figure \ref{fig:GPWilliamson}, we learn that Williamson tends to perform better in home matches, as indicated by the superior estimates for home games (represented by black bars) compared with away games (represented by orange bars).
Additionally, we see that Williamson tends to perform better in his team's first innings of a Test match, as indicated by the jagged behaviour in the estimates when observing changes between his first and second innings of the same match.

\begin{figure}[h]
	\centering
	\includegraphics[width = 0.7\linewidth]{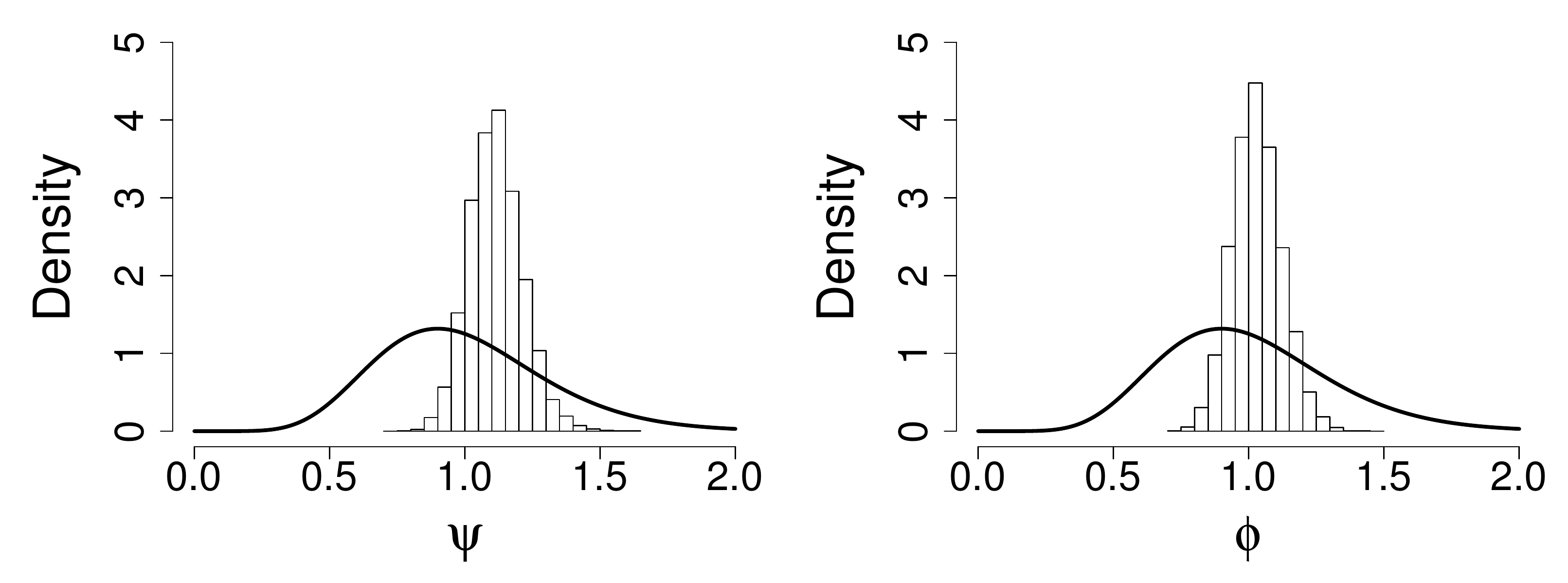}
	\caption{Parameter posterior distributions for $\psi$ and $\phi$ for Kane Williamson.
	Black lines indicate the prior distributions.
	The data provides moderate evidence that Williamson bats better at home and in New Zealand's first batting innings of the match.}
	\label{fig:match_posteriors}
\end{figure}

The posterior parameter summaries and posterior distributions summarised in Table \ref{tab:parameters} and Figure \ref{fig:match_posteriors} respectively, allow us to quantify the magnitude of these effects.
The multiplicative effect of playing at a home venue on runs scored (compared with a neutral venue) has a posterior mean of $\psi = 1.11$ and a 95\% credible interval of (0.93, 1.32).
We can obtain the estimates for comparing performance at a home venue, compared with an away venue, by squaring the parameter, giving a posterior mean $\psi^2 = 1.25$ and 95\% credible interval (0.87, 1.75).
Meanwhile, the multiplicative effect of batting in his team's first innings of a Test, compared with batting in his team's second innings, can also be obtained by squaring, giving a posterior mean $\phi^2 = 1.07$ and 95\% credible interval (0.74, 1.48).
\begin{table}[h]
	\caption{Posterior parameter summaries for Kane Williamson.}
	\centering
	\def\arraystretch{1.25} 
	\begin{tabular}{l c r r}
	\hline
	\textbf{Parameter} & \textbf{Mean} & \textbf{68\% C.I.} & \textbf{95\% C.I.} \\
	\hline
	$C$ & 0.30 & (0.23, 0.39) & (0.15, 0.53) \\
	$D$ & 0.12 & (0.06, 0.18) & (0.03, 0.29) \\
	$\lambda$ & 56.6 & (45.0, 67.5) & (31.3, 82.2) \\
	$\sigma$ & 0.27 & (0.11, 0.42) & (0.04, 0.74) \\
	$\ell$ & 36.7 & (12.1, 57.5) & (4.6, 129.8) \\
	$\alpha$ & 1.50 & (1.15, 1.84) & (1.02, 1.98) \\
	$\psi$ & 1.11 & (1.02, 1.21) & (0.93, 1.32) \\
	$\phi$ & 1.03 & (0.94, 1.12) & (0.86, 1.22) \\
	\hline
	\end{tabular}
	\label{tab:parameters}
\end{table}

These point estimates suggest that we should expect Williamson to score 25\% more runs at home venues, compared with away venues, and 7\% more runs in his team's first innings of a match, compared with the second innings.
However, given the reasonably large uncertainties for our estimates of $\psi$ and particularly $\phi$, there is no definitive evidence to say that Williamson necessarily bats better in home conditions, in his team's first innings of a Test match.

Across the group of 1018 players analysed, there was generally little evidence to support the presence of a venue or innings effect for players who have batted in fewer than 20 Test innings.
However, of the 474 players who have batted in 20 or more innings, the 68\% credible intervals provide evidence to suggest of a venue effect for 146 players and an innings effect for 168 players.

\begin{figure}[h]
	\centering
	\includegraphics[width = 0.7\linewidth]{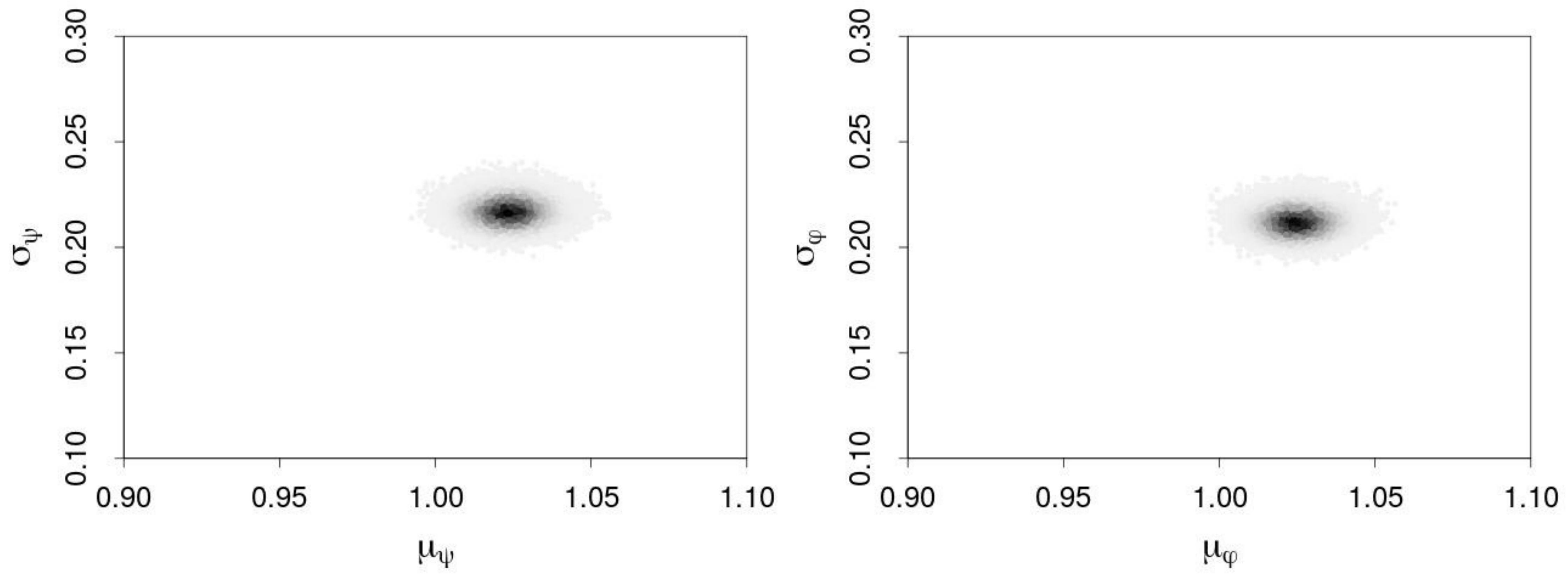}
	\caption{Joint posterior distributions for $\mu_\psi$ and $\sigma_\psi$, and $\mu_\phi$ and $\sigma_\phi$, shown across the uniform prior distributions for $\mu_\psi, \sigma_\psi, \mu_\phi$, and $\sigma_\phi$.
	Darker areas represent regions with higher posterior probability.}
	\label{fig:hierarchical_psi_phi}
\end{figure}

In order to generalise our inference regarding the venue and innings- specific effects across all players, we performed a hierarchical analysis for the relevant parameters, $\psi$ and $\phi$.
If we define a set of hyperparameters, $\{ \mu_\psi, \sigma_\psi, \mu_\phi, \sigma_\phi \}$, and implement a hierarchical model structure, we can quantify the typical values for $\psi$ and $\phi$ that players are clustered around, without having to analyse all the data jointly.
This is achieved by obtaining posterior estimates for $\psi$ and $\phi$, for each player, then post-processing the results to construct what the hierarchical model would have produced using MCMC samples \citep{hastings1970}.
Our hierarchical model takes the form:
\begin{align*}
	\mu_\psi, \mu_\phi & \sim \text{Uniform}(0.9, 1.1) \\
	\sigma_\psi, \sigma_\phi & \sim \text{Uniform}(0.1, 0.3) \\
	\text{log}(\psi) & \sim \text{Normal}(\textup{log}(\mu_\psi), \sigma_\psi^2) \\
	\text{log}(\phi) & \sim \text{Normal}(\textup{log}(\mu_\phi), \sigma_\phi^2)
\end{align*}
The joint posterior distributions for the set of hyperparameters, $\{ \mu_\psi, \sigma_\psi, \mu_\phi, \sigma_\phi \}$, are shown in Figure \ref{fig:hierarchical_psi_phi}.
The bulk of the posterior mass for both $\mu_\psi$ and $\mu_\phi$ are centered on values greater than 1, confirming our suspicion that the typical player performs better in home matches, in their team's first innings of a Test match.

\subsubsection{Quantifying career progression}
\begin{figure}[h]
	\centering
	\includegraphics[width = 0.7\linewidth]{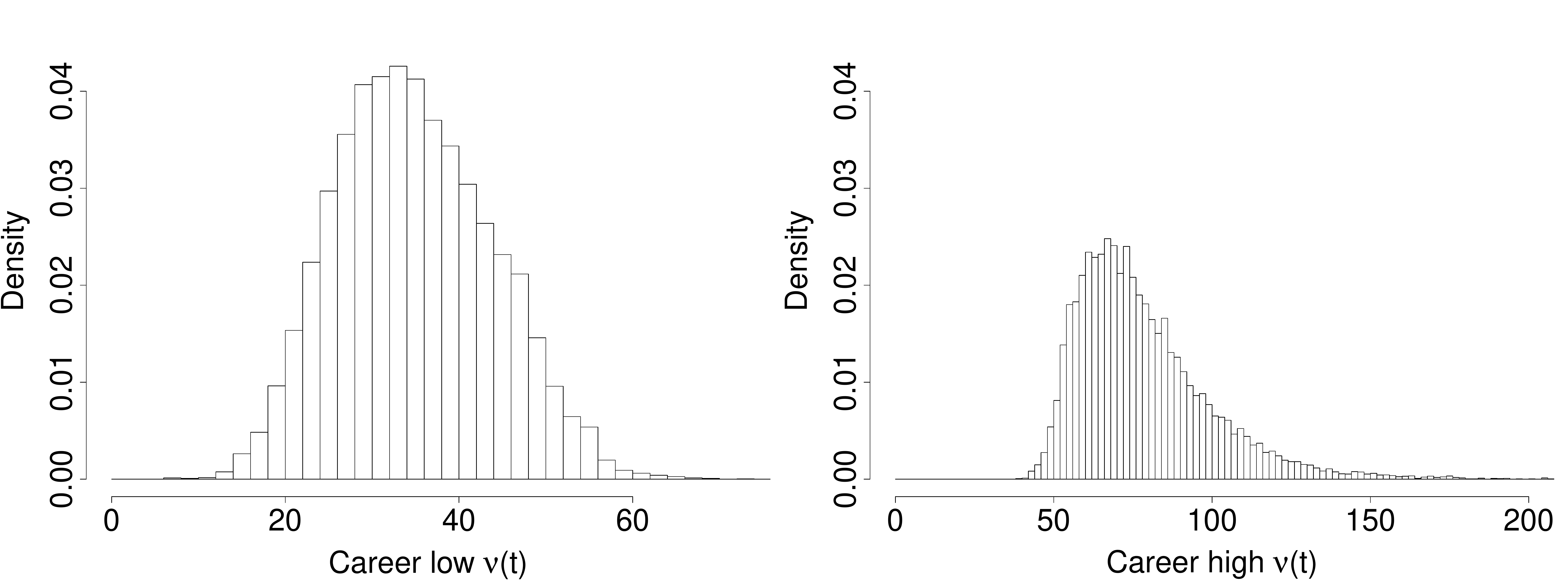}
	\caption{Posterior distributions for $\nu(t)$ at the lowest and highest points of Williamson's career.}
	\label{fig:minmax}
\end{figure}

\begin{figure}[H]
	\centering
	\includegraphics[width = 0.7\linewidth]{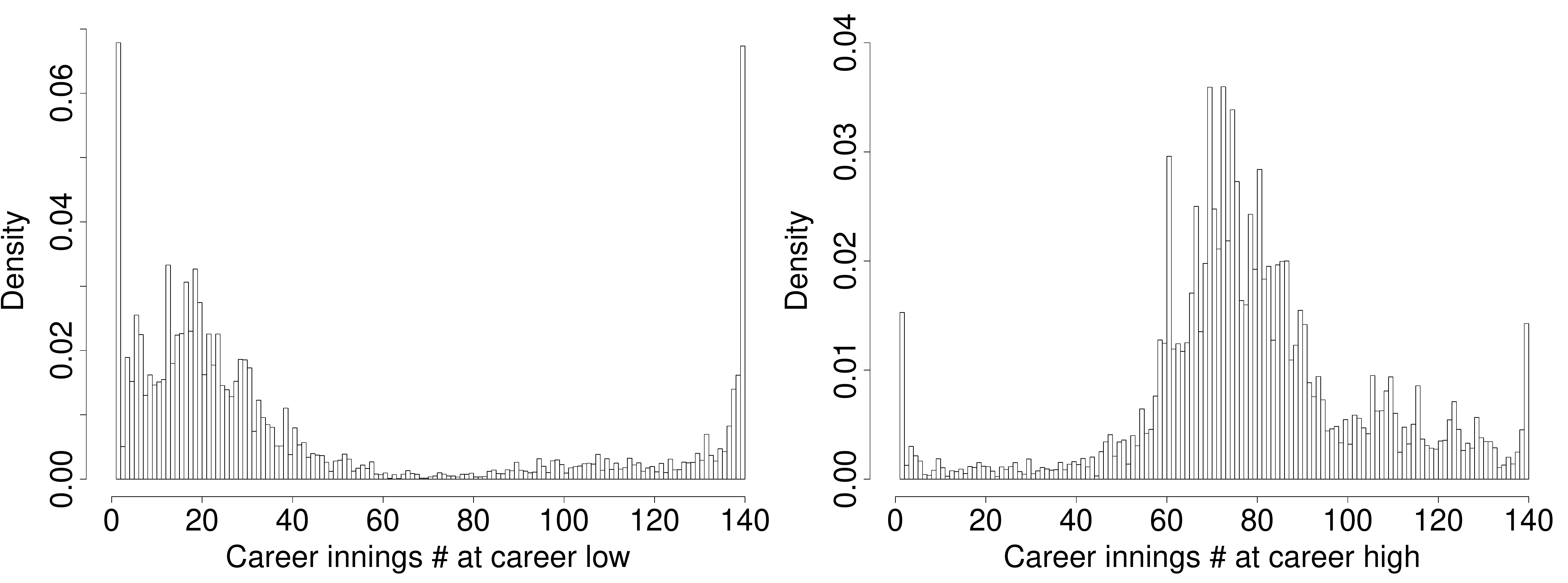}
	\caption{Posterior distributions for career innings index (time) of the lowest and highest points of Williamson's career.}
	\label{fig:minmax_index}
\end{figure}

Similar to the approach used in \cite{boys2019}, the model output also allows us to quantify a player's batting ability, in terms of a batting average, at their `lowest' and `highest' points of their career to date.
We are also able to predict the number of runs we expect them to score in their next career innings.
Each of these estimates assumes a neutral venue and that the innings number in the match is unknown.
The posterior distributions for $\nu(t)$ at Williamson's lowest and highest points of his career are shown in Figure \ref{fig:minmax} and are summarised in Table \ref{tab:williamson_highlow}.
Additionally, the posterior distributions for \emph{when} Williamson experienced his highest and lowest points of his career are shown in Figure \ref{fig:minmax_index}.
\begin{table}[h]
	\caption{Posterior point estimates for $\nu(t)$ at Williamson's lowest and highest points of his career and prediction for next career innings, including 68\% and 95\% credible intervals.}
	\centering
	\def\arraystretch{1.25}
	\begin{tabular}{l c r r}
		\hline
			& \textbf{Posterior median} & \textbf{68\% C.I.} & \textbf{95\% C.I.} \\
		\hline
		Career low $\nu(t)$ & 34.2 & (25.6, 44.5) & (19.0, 53.0) \\
		Career high $\nu(t)$ & 73.6 & (59.0, 97.2) & (50.5, 131.5) \\
		$\nu(t = \text{next innings})$ & 47.1 & (36.2, 57.9) & (25.3, 72.0) \\
		\hline
	\end{tabular}
	\label{tab:williamson_highlow}
\end{table}

For Williamson, we estimate that his underlying ability corresponded to an expected average of 34.2 at the lowest point of his career, while at his highest point, his expected average was 73.6.
Unsurprisingly, many of the posterior samples estimate Williamson's underlying ability was at its lowest near the beginning of his career, as suggested in Figure \ref{fig:minmax_index}.
Additionally, Figure \ref{fig:minmax_index} also suggests that Williamson most likely experienced his career peak to date sometime after his $60^\text{th}$ career innings.
The posterior distributions in Figure \ref{fig:minmax_index} generally support the idea of `finding your feet' for Williamson, as his underlying ability was very likely at its lowest near the beginning of his career, while his ability likely peaked only after playing in reasonable number of innings (or is still yet to peak).
Given Williamson has only recently turned 30 years of age, it is likely he still has a number of years left in his career where his underlying batting ability will be close to its best.

\subsection{Comparison of batting career trajectories}
\begin{figure}[!ht]
	\centering
	\includegraphics[width = 0.7\linewidth]{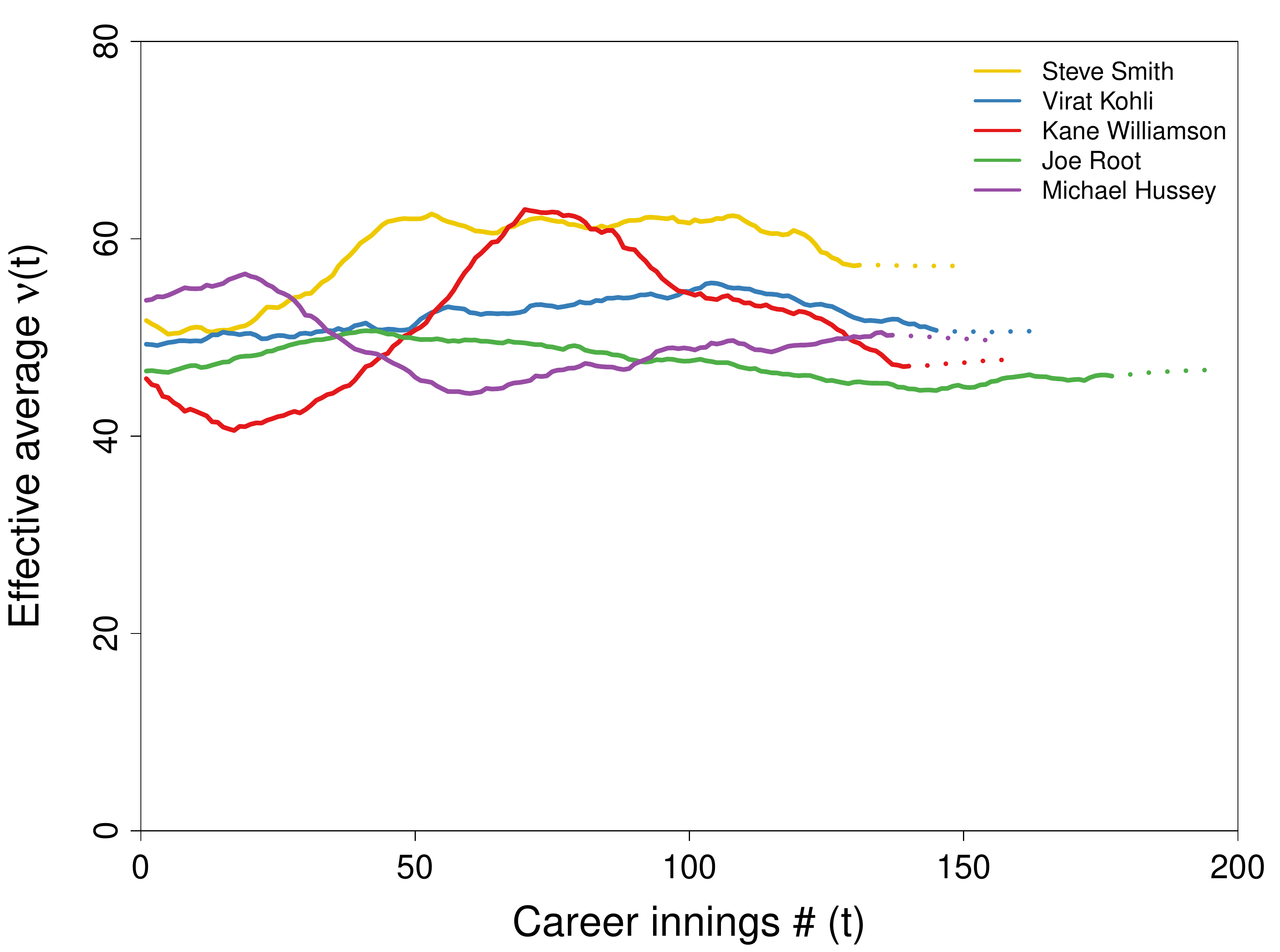}
	\caption{Test match batting career trajectories for the big four and Michael Hussey, including predictions for the next 20 innings (dotted).}
	\label{fig:GPall}
\end{figure}

In Figure \ref{fig:GPall}, we have plotted the career trajectories for `the big four' to illustrate how the model allows us to compare the career progressions of multiple batsmen.
All four players exhibit behaviour typical of finding your feet, taking a reasonable number of innings before reaching their peak ability, although different players appear to take different lengths of time to adjust to the demands of international cricket.
It is perhaps unsurprising to see Kane Williamson, as the player taking the longest time to fulfil his potential, as he was just 20 years old when making his Test match debut (compared with Smith (21), Kohli (22), and Root (21)).
As such, he was potentially a less developed batsman compared with the others at the time of his Test debut.

However, Figure \ref{fig:GPall} does suggest that Williamson is the player in the big four who has improved the most.
In the early stages of their respective careers, Williamson had the lowest estimated ability, however, the model also estimates him to have the had the highest ability at his peak.
On the contrary, current England Test captain, Joe Root, had a successful start to his career but his estimated batting ability in Test matches appears to have been on the decline for a number of innings.
This is not to say that Root is not a world-class batsman --- his predicted average of 46.2 for his next innings is impressive --- however, he has been a model of consistency, rather than exhibiting the same improvement in performance seen in Williamson, Smith, and to a lesser extent, Kohli.

The career trajectory for Australian batsman Michael Hussey is also presented in Figure \ref{fig:GPall}, as an example of a player who made their Test debut with a significant amount of domestic experience.
Hussey began his Test career at the age of 30 and appears to have immediately showcased his talent, not exhibiting the same typical finding your feet behaviour as many players who debut in their early twenties.
While no amount of domestic cricket can truly prepare a player for the trials and tribulations of Test cricket, such observations may provide credence to the idea that more experienced players making their Test debut may tend to require less time to reach their peak ability.

Therefore, given the large innings-to-innings variation in an individual's career scores (as observed in Figure \ref{fig:GPWilliamson}) and due to the idea of finding your feet (as observed in Figure \ref{fig:GPall}), it may be unwise to make a judgement regarding a young player's underlying batting ability after just several innings.
While it may not always be practical to afford every single batsman ample opportunity to reach their peak ability, particularly within the cut-throat confines of international cricket, it does at least suggest that there may be little to learn about a player's true skill-ceiling from just a few innings.

\subsection{Player rankings}
The current top 20 Test batsmen as estimated by the Gaussian process model are presented in Table \ref{tab:predictions}, where players are ranked by the expected number of runs we predict them to score in their next career innings, assuming a neutral venue and the innings number in the match is unknown (an up-to-date list of the top 100 players is maintained at \url{www.oliverstevenson.co.nz/\#research}).
For comparison, each player's ICC rating and world ranking (\#) are also provided.
The ranking of players is generally similar between the two methods, although there are several notable differences.
\begin{table}[h]
	\caption{Current (as of $1^\text{st}$ December 2020) top 20 Test match batsmen ranked by expected number of runs scored in their next career innings, $\nu(t = \text{next innings})$, including the 68\% credible interval. ICC Test batting ratings and world rankings (\#) are shown for comparison.}
	\centering
	\resizebox{\textwidth}{!}
   	{
	\def\arraystretch{1.2} 
	\begin{tabular}{l l c c r r}
		\hline
		\textbf{Rank} & \textbf{Player} & \textbf{Innings} & \textbf{Career average} & $\boldsymbol{\nu(t = \textbf{next innings})}$ & \textbf{ICC rating (\#)} \\
		\hline
		1. & S. Smith \hfill (AUS) & 131 & 62.8 & 57.9 (47.9, 68.8) & \hfill 911 \hfill (1) \\
		2. & M. Labuschagne \hfill (AUS) & 23 & 63.4 & 55.8 (43.0, 75.2) & \hfill 827 \hfill (3) \\
		3. & B. Azam \hfill (PAK) & 53 & 45.4 & 53.5 (40.8, 75.8) & \hfill 797 \hfill (5) \\
		4. & V. Kohli \hfill (IND) & 145 & 53.6 & 51.4 (43.4, 59.8) & \hfill 886 \hfill (2) \\
		5. & D. Warner \hfill (AUS) & 155 & 48.9 & 48.0 (41.1, 56.8) & \hfill 793 \hfill (6) \\
		6. & K. Williamson \hfill (NZ) & 140 & 51.0 & 47.7 (36.8, 58.9) & \hfill 812 \hfill (4) \\
		7. & A. Mathews \hfill (SL) & 154 & 45.3 & 47.6 (40.4, 59.3) & \hspace{0.7cm} 658 \hfill (17) \\
		8. & R. Sharma \hfill (IND) & 53 & 46.5 & 46.6 (37.7, 58.9) & \hspace{0.7cm} 674 \hfill (16) \\
		9. & M. Agarwal \hfill (IND) & 17 & 57.3 & 46.2 (34.2, 63.5) & \hspace{0.7cm} 714 \hfill (11) \\ 		
		10. & J. Root \hfill (ENG) & 177 & 48.0 & 45.7 (39.4, 52.3) & \hfill 738 \hfill (9) \\
		11. & R. Taylor \hfill (NZ) & 178 & 46.1 & 43.6 (36.9, 50.6) & \hspace{0.7cm} 677 \hfill (15) \\
		12. & C. Pujara \hfill (IND) & 128 & 48.7 & 43.4 (36.6, 50.6) & \hspace{0.7cm} 766 \hfill (6) \\
		13. & A. Ali \hfill (PAK) & 152 & 42.9 & 42.8 (35.5, 52.1) & \hspace{0.7cm} 627 \hfill (23) \\
		14. & A. Rahane \hfill (IND) & 109 & 42.9 & 41.6 (34.3, 50.5) & \hspace{0.7cm} 726 \hfill (10) \\
		15. & T. Latham \hfill (NZ) & 92 & 42.3 & 40.8 (33.9, 48.8) & \hspace{0.7cm} 710 \hfill (12) \\
		16. & M. Rahim \hfill (BAN) & 130 & 36.8 & 40.5 (33.8, 54.6) & \hspace{0.7cm} 654 \hfill (18) \\
		17. & B. Stokes \hfill (ENG) & 122 & 37.8 & 39.4 (33.3, 47.7) & \hspace{0.7cm} 760 \hfill (8) \\
		18. & D. Chandimal \hfill (SL) & 103 & 40.8 & 39.3 (31.3, 47.2) & \hspace{0.7cm} 563 \hfill (28) \\	
		19. & T. Head \hfill (AUS) & 28 & 42.0 & 38.1 (29.5, 49.2) & \hspace{0.7cm} 643 \hfill (20) \\
		20. & BJ. Watling \hfill (NZ) & 110 & 38.5 & 38.1 (31.1, 45.4) & \hspace{0.7cm} 621 \hfill (25) \\
		\hline
	\end{tabular}
	}
	\label{tab:predictions}
\end{table}

Firstly, the Gaussian process model ranks Indian batsman Rohit Sharma $8^\text{th}$ in the world, while he is ranked $16^\text{th}$ according to the current ICC ratings.
Upon closer inspection, it appears as though Sharma has a higher than usual proportion of not out scores that are greater than 50, suggesting he frequently overcomes the getting your eye in process, but for various reasons has not had the opportunity to convert these starts into big scores.
The Gaussian process model rewards players who are able to get their eye in and remain on a not out out score, while the ICC ratings simply provide not out innings with a `bonus' that we suspect is too low.
In this sense, we agree with \cite{boys2019}, that the ICC rating system does not appropriately adjust for not out scores in a manner that takes account of player ability.

Secondly, the ICC ranking method tends to place a greater emphasis on more recent innings, compared with the Gaussian process model.
For example, the current ICC ratings rank English all-rounder, Ben Stokes, as the $8^\text{th}$ best Test batsman globally, however, the Gaussian process model ranks Stokes at $17^\text{th}$.
In his last 10 Test innings, Stokes has averaged 58.0 runs per dismissal, well above his career average of 37.8, suggesting that the difference in rankings is likely due to the ICC method placing a greater emphasis on recent scores.
Conversely, the Gaussian process model is more conservative when estimating batting ability with respect to short-term form.
Unfortunately, due to the closed source nature of the ICC ratings formula, we do not know how much emphasis is placed on players' recent innings.
An exponential weighted average is used, but the scale length is not public knowledge. This makes it difficult to pinpoint the exact differences in the methods.

Finally, Sri Lankan batsmen Angelo Mathews and Dinesh Chandimal appear to be victims of ICC rating decay, whereby players who have not batted in a recent match, see their rating slowly decline.
While well intentioned to reflect the best current batsmen in the world, the decay system is inherently biased as it tends to affect smaller nations disproportionately, who are generally afforded fewer opportunities to play Test cricket.

Therefore, while both methods provide a general indication of batting ability, the Gaussian process model has the added ability of quantifying the differences in ability between players in a more meaningful manner.
For example, rather than concluding `Steve Smith is 99 rating points better than Kane Williamson', we can make more useful probabilistic statements by computing $P(\nu_{Smith}(t) > \nu_{Williamson}(t) \ | \ t = \text{next innings})$, such as, `we estimate Smith has a 56.5	\% chance of outscoring Williamson in their next respective innings or `we expect Smith to outscore Williamson in their next respective innings by 10.2 runs', assuming a neutral venue and it is unknown whether they are batting in their teams' first or second innings.
The venue and innings-specific effects can easily be taken into account if these variables are known for each players' next innings.
It is also worth noting that when computing these probabilities, we account for the getting your eye in process, as these predictions are all based on the posterior predictive distribution under our model.

Although the ICC ratings account for more variables, such as opposition strength and pitch effects, it is far more intuitive to quantify differences in batting ability between players by using a natural cricketing interpretation, such as units of a batting average, rather than arbitrary rating points.
In a practical sense, the results from the Gaussian process model has significant implications in the likes of player comparison and team selection policy, as coaches and selectors will have a more direct means of understanding the risks and real life impacts of selecting one player over another.

\section{Model prediction and comparison}
\subsection{Model prediction} \label{sec:error}
When it comes to prediction, our primary goal is to be able to accurately predict the \emph{next} score in a batsman's career, while our secondary goal is to best explain how a player's ability has fluctuated over the course of their career to date.
Therefore, when we assess the predictive capabilities of the Gaussian process model, we compute the model predictions using leave-one-out cross-validation \citep{sammut2010}, whereby we leave out the player's most recent innings, obtain a prediction for the score in this innings, and compare the model prediction with the actual observed score.
In order to avoid the complexities that arise with not out innings, we have opted to predict the most recent `out' score for each player when computing the prediction errors.

Prediction errors have been computed for the Gaussian process model, as well as a range of `simple moving average' (SMA) models of different orders for comparison.
The SMA models compute a prediction for a player's next career score, based on the previous 10\%, 25\%, 50\%, and 100\% of a player's career innings.
For example, if a player has batted in 100 career innings, the SMA(10\%) model will use the most recent 10\% of the player's career data (in this case, their last 10 innings) to predict their next innings score, while the SMA(50\%) model would make a prediction using the most recent 50\% of the player's career data (their last 50 innings).
Note that the SMA(100\%) model is equivalent to predicting batting ability using the career batting average and assumes a player's ability is constant throughout their career.

We can then compare the predictive performance of the Gaussian process model with the SMA models by comparing the respective model predictions and prediction errors.
As leave-one-out cross-validation requires us to leave out the players most recent out innings, we are unable to compute the prediction errors for players who have only batted in one innings, or for players who have never been dismissed during their Test career.
Of the 1018 players analysed, we were able to compute the prediction errors using all five models for 913 players.

\begin{table}[h]
	\caption{Mean squared prediction errors (MSE) using leave-one-out cross-validation.
	The Gaussian process model outperforms all SMA models across all players, while the SMA(10\%) model tends to perform worst of all.}
	\centering
	\resizebox{\textwidth}{!}
   	{
	\def\arraystretch{1.25} 
	\begin{tabular}{l r r r r}
		\hline
		& \multicolumn{4}{c}{\textbf{Minimum \# of career innings}} \\
		\textbf{Model} & \textbf{No minimum} & \textbf{10 innings} & \textbf{20 innings} & \textbf{50 innings} \\
		\hline
		SMA(10\%) model & 633.1 & 751.3 & 684.2 & 857.7 \\
		SMA(25\%) model & 588.4 & 696.3 & 646.1 & 837.7 \\
		SMA(50\%) model & 608.2 & 704.2 & 661.7 & 859.0 \\
		SMA(100\%) model & 589.1 & 681.9 & 655.6 & 829.3 \\
		Gaussian process model & 544.0 & 649.6 & 616.8 & 785.8 \\
		\hline
	\end{tabular}
	}
	\label{tab:prediction}
\end{table}

As indicated in Table \ref{tab:prediction}, the Gaussian process model outperforms all other models in terms of prediction, averaged across all players.
Additionally, unlike the SMA models, the Gaussian process model is able to deal with not out scores, predict future estimates of batting ability, and do a far better job of explaining a player's career trajectory to date, as the SMA model estimates tend to be rather erratic in this regard.
Interestingly, the SMA(10\%) model tends to perform worst of all, suggesting that predicting a player's ability purely on recent scores is unwise.

\subsection{Model comparison}
We are able to compare the fit of individual models directly using the marginal likelihood or `evidence', $Z$.
For a Bayesian model with parameters $\theta$ and data $d$, the evidence of a model $M$, is the prior probability of the data given the model, and can be computed using:
\begin{equation}\label{eq:evidence}
	Z_M = P(d \,|\, M) = \int P(\theta \,|\, M) \ P(d \,|\, \theta, M) \ d\theta.
\end{equation}

A major advantage of using nested sampling to fit each model is that the evidence is computed in addition to posterior samples, making for trivial model comparison \citep{skilling2006}.
Here, we use the evidence to compare the support for the Gaussian process model, $Z$, against the SMA(100\%) model that assumes a player's ability remains constant throughout their career, $Z_0$.
The logarithm of the Bayes factor between the two models are presented in Table \ref{tab:evidence}, for our top 10 ranked players, while the sum of the model likelihoods for all players are shown in the bottom row, alongside the logarithm of the Bayes factor, averaged across all players.
\begin{table}[h]
	\caption{Evidence for the Gaussian process and SMA(100\%) models.
	The right hand column, the logarithm of the Bayes factor, shows the data generally support the Gaussian process model over the SMA(100\%) model.}
	\centering
	\def\arraystretch{1.25}
	\begin{tabular}{l l r r r r}
		\hline
		\textbf{Rank} & \textbf{Player} & log($Z$) & log($Z_0$) & log($\frac{Z}{Z_0}$) \\
		\hline
		1. & S. Smith \hfill (AUS) & $-592.0$ & $595.0$ & 3.0 \\
		2. & M. Labuschange \hfill (AUS) & $-120.5$ & $-120.7$ & 0.2 \\
		3. & B. Azam \hfill (PAK) & $-214.7$ & $-219.1$ & 4.4 \\
		4. & V. Kohli \hfill (IND) & $-669.6$ & $-676.5$ & 6.9 \\
		5. & D. Warner \hfill (AUS) & $-721.9$ & $-727.9$ & 6.0 \\
		6. & K. Williamson \hfill (NZ) & $-623.9$ & $-630.2$ & 6.3 \\
		7. & A. Mathews \hfill (SL) & $-642.4$ & $-639.3$ & -3.1 \\
		8. & R. Sharma \hfill (IND) & $-219.3$ & $-225.1$ & 5.8 \\
		9. & M. Agarwal \hfill (IND) & $-86.8$ & $-87.7$ & 0.9 \\
		10. & J. Root \hfill (ENG) & $-797.9$ & $-798.3$ & 0.4 \\
		\hline
		& \textbf{All players} & $-152071.8$ & $-153473.4$ & $1.4$ \\
		\hline
	\end{tabular}
	\label{tab:evidence}
\end{table}

Overall, for most players the Gaussian process model is preferred over the SMA(100\%) model that assumes a constant batting ability.
Generally speaking, the more innings a player has batted in, the more likely the Gaussian process model will be the preferred model.
As our nested sampling scheme is inherently a Monte Carlo process, the estimates for the evidence are not exact.
However, the algorithm was run with a large number of particles and MCMC iterations, ensuring our Monte-Carlo related errors are negligible.
Assuming efficient MCMC exploration for generating the new points, the standard errors on the $\log(Z)$ estimates due to Monte Carlo are roughly $\pm 0.1$ for each player.

\section{Concluding remarks and further work}

We have presented a novel method of estimating the underlying batting abilities of individual cricket players, and have demonstrated how ability can vary and fluctuate over the course of a playing career.
The results provide a more accurate means of quantifying batting ability at any given point of a player's career, compared with traditional cricket metrics, such as the batting average.
We have also found evidence to support several of our pre-conceived hunches, namely that the majority of batsmen score more runs when batting in their team's first innings of a Test match, at a home venue.
Additionally, the findings generally support the cricketing concept of finding your feet, whereby players do not begin their careers batting to the best of their ability.
Instead, many players appear to take a number of innings to adjust to the demands of international cricket, before reaching their peak batting ability.

For the majority of players analysed, little evidence was found to support the claim that recent, short-term form has a significant impact when it comes to estimating a player's current batting ability.
Rather, estimates of player batting ability tend to change slowly over time, suggesting a player's underlying ability is more likely to be affected gradually due to the likes of age, experience and general changes in technique and fitness, rather than jumping erratically from innings-to-innings as a result of recent performances.
The findings suggest the effect of recent form should be considered on an individual basis, with some players more influenced by recent performances than others.
Often, both domestic and international fans will cite recent good form as a reason to select a particular player, or will use a recent run of poor form as an excuse to drop an incumbent.
Of course, if a player's underlying ability appears to have been on the decline for some time, then there may be a valid reason to consider swinging the selection axe.
However, our findings would suggest that selecting players purely on the basis of recent form may be a classic case of falling victim to recency bias, especially if said players do not have a strong career record to fall back upon.

We have also compared the rankings of current Test batsman using the proposed Gaussian process model, with the more established ICC Test batting rankings and found there to be a reasonable amount of overlap between the two methods.
However, the present model has the added advantage of maintaining an intuitive cricketing interpretation by measuring ability in units of a batting average, as opposed to a rating with no meaningful interpretation.
This allows for the results of the model to be more easily digested by the likes of coaches and selectors who may lack formal statistical training, but have plenty of experience with batting averages.
Furthermore, as the model allows us to make probabilistic statements when comparing the abilities of multiple players, we are easily able to predict and quantify the risks and rewards of selecting one player over another.
Consequently, the findings may have practical implications in terms of talent identification and team selection policy.

It is worth acknowledging that we have ignored several important variables, such as the number of balls faced in each innings and the strength of the opposition.
Presently, the model treats all runs scored equally, but, finding a way of rewarding batsmen for consistently scoring runs against world-class bowling attacks will further enhance the predictive capabilities of the model.

The natural next step in this field would be to apply the model to the performance of bowlers over time, as well as widening the scope of the model to include one-day and even Twenty20 matches.
As the present model provides us with estimates of individual batting ability at \emph{every} point of a player's career, any model that is developed to evaluate bowlers can use these estimates to then account for the strength of the batsmen to whom they bowled.

\section*{Acknowledgements}
This research is funded by a University of Auckland Doctoral scholarship.
The authors wish to acknowledge the use of \href{https://www.nesi.org.nz}{New Zealand eScience Infrastructure (NeSI)} high performance computing facilities, consulting support and/or training services as part of this research. New Zealand's national facilities are provided by NeSI and funded jointly by NeSI's collaborator institutions and through the Ministry of Business, Innovation \& Employment's Research Infrastructure programme.

It is also a pleasure to thank Pete Mayell (NV Play), Matt Smith (NV Play), and Gus Pickering (NV Play) for their helpful discussions.
We would also like to thank the reviewers and journal editors for their constructive comments and suggestions that helped improve the paper.

\bibliographystyle{apalike}
\bibliography{references}

\begin{thebibliography}{}

\bibitem[Bailey and Clarke, 2006]{bailey2006}
Bailey, M. and Clarke, S.~R. (2006).
\newblock Predicting the match outcome in one day international cricket
  matches, while the game is in progress.
\newblock {\em Journal of Sports Science \& Medicine}, 5(4):480.

\bibitem[Boys and Philipson, 2019]{boys2019}
Boys, R.~J. and Philipson, P.~M. (2019).
\newblock On the ranking of test match batsmen.
\newblock {\em Journal of the Royal Statistical Society: Series C (Applied
  Statistics)}, 68(1):161--179.

\bibitem[Bracewell and Ruggiero, 2009]{bracewell2009}
Bracewell, P.~J. and Ruggiero, K. (2009).
\newblock A parametric control chart for monitoring individual batting
  performances in cricket.
\newblock {\em Journal of Quantitative Analysis in Sports}, 5(3).

\bibitem[Brewer, 2008]{brewer2008}
Brewer, B.~J. (2008).
\newblock Getting your eye in: a {B}ayesian analysis of early dismissals in
  cricket.
\newblock {\em arXiv preprint arXiv:0801.4408}.

\bibitem[Brooker and Hogan, 2011]{brooker2011}
Brooker, S. and Hogan, S. (2011).
\newblock A method for inferring batting conditions in {ODI} cricket from
  historical data.

\bibitem[Brooks et~al., 2002]{brooks2002}
Brooks, R.~D., Faff, R.~W., and Sokulsky, D. (2002).
\newblock An ordered response model of {T}est cricket performance.
\newblock {\em Applied Economics}, 34(18):2353--2365.

\bibitem[Cai et~al., 2002]{cai2002}
Cai, T., Hyndman, R.~J., and Wand, M. (2002).
\newblock Mixed model-based hazard estimation.
\newblock {\em Journal of Computational and Graphical Statistics},
  11(4):784--798.

\bibitem[Carter and Guthrie, 2004]{carter2004}
Carter, M. and Guthrie, G. (2004).
\newblock Cricket interruptus: fairness and incentive in limited overs cricket
  matches.
\newblock {\em Journal of the Operational Research Society}, 55(8):822--829.

\bibitem[Clarke and Norman, 1999]{clarke1999}
Clarke, S.~R. and Norman, J.~M. (1999).
\newblock To run or not?: Some dynamic programming models in cricket.
\newblock {\em Journal of the Operational Research Society}, 50(5):536--545.

\bibitem[Clarke and Norman, 2003]{clarke2003}
Clarke, S.~R. and Norman, J.~M. (2003).
\newblock Dynamic programming in cricket: Choosing a night watchman.
\newblock {\em Journal of the Operational Research Society}, 54(8):838--845.

\bibitem[Clarke et~al., 1998]{clarke1998}
Clarke, S.~R., Norman, J.~M., et~al. (1998).
\newblock Dynamic programming in cricket: protecting the weaker batsman.
\newblock {\em Asia Pacific Journal of Operational Research}, 15:93--108.

\bibitem[Csapo et~al., 2015]{csapo2015}
Csapo, P., Avugos, S., Raab, M., and Bar-Eli, M. (2015).
\newblock The effect of perceived streakiness on the shot-taking behaviour of
  basketball players.
\newblock {\em European Journal of Sport Science}, 15(7):647--654.

\bibitem[Davis et~al., 2015]{davis2015}
Davis, J., Perera, H., and Swartz, T.~B. (2015).
\newblock A simulator for {T}wenty20 cricket.
\newblock {\em Australian \& New Zealand Journal of Statistics}, 57(1):55--71.

\bibitem[Duckworth and Lewis, 1998]{duckworthlewis1998}
Duckworth, F.~C. and Lewis, A.~J. (1998).
\newblock A fair method for resetting the target in interrupted one-day cricket
  matches.
\newblock {\em Journal of the Operational Research Society}, 49(3):220--227.

\bibitem[Durbach and Thiart, 2007]{durbach2007}
Durbach, I.~N. and Thiart, J. (2007).
\newblock On a common perception of a random sequence in cricket: application.
\newblock {\em South African Statistical Journal}, 41(2):161--187.

\bibitem[Elderton and Wood, 1945]{elderton1945}
Elderton, W. and Wood, G.~H. (1945).
\newblock Cricket scores and geometrical progression.
\newblock {\em Journal of the Royal Statistical Society}, 108(1/2):12--40.

\bibitem[Gilovich et~al., 1985]{gilovich1985}
Gilovich, T., Vallone, R., and Tversky, A. (1985).
\newblock The hot hand in basketball: On the misperception of random sequences.
\newblock {\em Cognitive psychology}, 17(3):295--314.

\bibitem[Hastings, 1970]{hastings1970}
Hastings, W.~K. (1970).
\newblock Monte carlo sampling methods using markov chains and their
  applications.

\bibitem[Ian and Thomas, 2002]{ian2002}
Ian, P. and Thomas, J. (2002).
\newblock Rain rules for limited overs cricket and probabilities of victory.
\newblock {\em Journal of the Royal Statistical Society: Series D (The
  Statistician)}, 51(2):189--202.

\bibitem[Jayadevan, 2002]{jayadevan2002}
Jayadevan, V. (2002).
\newblock A new method for the computation of target scores in interrupted,
  limited-over cricket matches.
\newblock {\em Current Science}, 83(5):577--586.

\bibitem[Kimber and Hansford, 1993]{kimber1993}
Kimber, A.~C. and Hansford, A.~R. (1993).
\newblock A statistical analysis of batting in cricket.
\newblock {\em Journal of the Royal Statistical Society. Series A (Statistics
  in Society)}, pages 443--455.

\bibitem[Koulis et~al., 2014]{koulis2014}
Koulis, T., Muthukumarana, S., and Briercliffe, C.~D. (2014).
\newblock A {B}ayesian stochastic model for batting performance evaluation in
  one-day cricket.
\newblock {\em Journal of Quantitative Analysis in Sports}, 10(1):1--13.

\bibitem[Lewis, 2003]{lewis2004moneyball}
Lewis, M. (2003).
\newblock {\em Moneyball: The art of winning an unfair game}.
\newblock WW Norton \& Company.

\bibitem[MacKay, 2003]{mackay2003}
MacKay, D.~J. (2003).
\newblock {\em Information theory, inference and learning algorithms}.
\newblock Cambridge university press.

\bibitem[Morley and Thomas, 2005]{morley2005}
Morley, B. and Thomas, D. (2005).
\newblock An investigation of home advantage and other factors affecting
  outcomes in english one-day cricket matches.
\newblock {\em Journal of sports sciences}, 23(3):261--268.

\bibitem[Nevill and Holder, 1999]{nevill1999}
Nevill, A.~M. and Holder, R.~L. (1999).
\newblock Home advantage in sport.
\newblock {\em Sports Medicine}, 28(4):221--236.

\bibitem[Norman and Clarke, 2010]{norman2010}
Norman, J.~M. and Clarke, S.~R. (2010).
\newblock Optimal batting orders in cricket.
\newblock {\em Journal of the Operational Research Society}, 61(6):980--986.

\bibitem[Pollard, 1986]{pollard1986}
Pollard, R. (1986).
\newblock Home advantage in soccer: A retrospective analysis.
\newblock {\em Journal of sports sciences}, 4(3):237--248.

\bibitem[Preston and Thomas, 2000]{preston2000}
Preston, I. and Thomas, J. (2000).
\newblock Batting strategy in limited overs cricket.
\newblock {\em Journal of the Royal Statistical Society: Series D (The
  Statistician)}, 49(1):95--106.

\bibitem[Rasmussen and Williams, 2006]{rasmussenwilliams2006}
Rasmussen, C.~E. and Williams, C. K.~I. (2006).
\newblock {\em Gaussian processes for machine learning}.
\newblock MIT Press.

\bibitem[Sahni and Bhogal, 2017]{sahni2017}
Sahni, M. and Bhogal, G. (2017).
\newblock Anxiety, depression and perceived sporting performance among
  professional cricket players.
\newblock {\em Br J Sports Med}, pages bjsports--2017.

\bibitem[Sammut and Webb, 2010]{sammut2010}
Sammut, C. and Webb, G.~I. (2010).
\newblock Leave-one-out cross-validation.
\newblock {\em Encyclopedia of Machine Learning}, pages 600--601.

\bibitem[Santos-Fernandez et~al., 2019]{santos2019}
Santos-Fernandez, E., Wu, P., and Mengersen, K.~L. (2019).
\newblock Bayesian statistics meets sports: A comprehensive review.
\newblock {\em Journal of Quantitative Analysis in Sports}.

\bibitem[Scarf et~al., 2011]{scarf2011}
Scarf, P., Shi, X., and Akhtar, S. (2011).
\newblock On the distribution of runs scored and batting strategy in test
  cricket.
\newblock {\em Journal of the Royal Statistical Society: Series A (Statistics
  in Society)}, 174(2):471--497.

\bibitem[Skilling, 2006]{skilling2006}
Skilling, J. (2006).
\newblock Nested sampling for general {B}ayesian computation.
\newblock {\em Bayesian analysis}, 1(4):833--859.

\bibitem[Stern, 2016]{stern2016duckworth}
Stern, S.~E. (2016).
\newblock The {D}uckworth-{L}ewis-{S}tern method: extending the
  {D}uckworth-{L}ewis methodology to deal with modern scoring rates.
\newblock {\em Journal of the Operational Research Society}, 67(12):1469--1480.

\bibitem[Stevenson, 2017]{stevenson2017masters}
Stevenson, O.~G. (2017).
\newblock The {N}ervous 90s: {A} {B}ayesian {A}nalysis of {B}atting in {T}est
  {C}ricket.
\newblock Master's thesis, University of Auckland.

\bibitem[Stevenson and Brewer, 2017]{stevenson2017}
Stevenson, O.~G. and Brewer, B.~J. (2017).
\newblock Bayesian survival analysis of batsmen in {T}est cricket.
\newblock {\em Journal of Quantitative Analysis in Sports}, 13(1):25--36.

\bibitem[Stevenson and Brewer, 2018]{stevenson2018}
Stevenson, O.~G. and Brewer, B.~J. (2018).
\newblock Modelling career trajectories of cricket players using {G}aussian
  processes.
\newblock In {\em Bayesian Statistics and New Generations}, pages 165--173.
  Springer.

\bibitem[Stroustrup, 2013]{c++}
Stroustrup, B. (2013).
\newblock {\em The C++ programming language}.
\newblock Pearson Education.

\bibitem[Swartz et~al., 2006]{swartz2006}
Swartz, T.~B., Gill, P.~S., Beaudoin, D., et~al. (2006).
\newblock Optimal batting orders in one-day cricket.
\newblock {\em Computers \& operations research}, 33(7):1939--1950.

\bibitem[Swartz et~al., 2009]{swartz2009}
Swartz, T.~B., Gill, P.~S., and Muthukumarana, S. (2009).
\newblock Modelling and simulation for one-day cricket.
\newblock {\em Canadian Journal of Statistics}, 37(2):143--160.

\bibitem[Totterdell, 1999]{totterdell1999}
Totterdell, P. (1999).
\newblock Mood scores: Mood and performance in professional cricketers.
\newblock {\em British Journal of Psychology}, 90(3):317--332.

\bibitem[Tversky and Gilovich, 1989]{tversky1989}
Tversky, A. and Gilovich, T. (1989).
\newblock The cold facts about the “hot hand” in basketball.
\newblock {\em Chance}, 2(1):16--21.

\end{thebibliography}

\end{document}